\documentclass[prd,aps,eqsecnum,floats,twocolumn,amsfonts]{revtex4}
\usepackage{epsfig}
\usepackage{hyperref}
\begin{document}

  \title{A $\chi^2$ time--frequency discriminator for gravitational wave detection}
  
  \author{Bruce Allen} 
  \email{ballen@uwm.edu} 
  \affiliation{Department of Physics, University of
    Wisconsin - Milwaukee, P.O. Box 413, Milwaukee WI 53201}
  
  \date{Draft of May 28, 2004}
    
  \begin{abstract}
    Searches for known waveforms in gravitational wave detector data
    are often done using matched filtering.  When used on real
    instrumental data, matched filtering often does not perform as
    well as might be expected, because non-stationary and non-Gaussian
    detector noise produces large spurious filter outputs (events).
    This paper describes a $\chi^2$ time-frequency test which is one
    way to discriminate such spurious events from the events that
    would be produced by genuine signals. The method works well only
    for broad-band signals. The case where the filter template does
    not exactly match the signal waveform is also considered, and
    upper bounds are found for the expected value of $\chi^2$.
  \end{abstract}

  \maketitle

\section{INTRODUCTION}
\label{s:intro}

Matched filtering is a common and effective technique used to search
for signals with a known waveform in a data stream \cite{Helstrom}.
The output of a matched filter will be large if the data stream
contains the desired signal.  But it can also be driven to large
values by spurious noise.  This paper describes a $\chi^2$
time-frequency discriminator statistic which has proven effective at
distinguishing between these two possibilities.

The method was invented by the author in 1996 \cite{grasp_chi2} for
use on data from the LIGO 40m prototype gravitational wave detector
\cite{40inspiral}.  The method was subsequently used in the analysis
of data from the Japanese TAMA detector \cite{tamainspiral1,
tamainspiral2, tamainspiral3, tamainspiral4, tamainspiral5} and the
first analysis of science data from the full-scale LIGO detectors
\cite{ligo_inspiral}.  It has also been used in preliminary searches
using VIRGO engineering data \cite{virgoinspiral} and GEO-600 data
\cite{Babak}.  Until now the only detailed description of the method
was in the documentation for the GRASP software package
\cite{grasp_chi2}, as referenced in the publications above. This paper
describes the method in more detail, and analyzes its properties.

The $\chi^2$ time-frequency discriminator is designed for use with
broad-band signals and detectors. The essence of the test is to
``break up'' the instrument's bandwidth into several smaller bands,
and to see if the response in each band is consistent with what would
be expected from the purported signal.  This method can only be used
to discriminate signals for which the gravitational waveform is {\it
known}, meaning that it can be calculated in advance, with high
precision.  (Note: the word ``known'' is slightly misleading since the
waveform typically still depends upon a few unknown parameters, such
as the overall scale and initial phase.)

A new generation of broadband gravitational wave detectors is now
undergoing commissioning \cite{virgo} and more sensitive instruments
are in the planning and design stages \cite{tama2,ligo2}.  We expect
that this test will prove useful for those instruments as well.

In searching for signals and setting upper limits, the primary use of
the $\chi^2$ time-frequency discriminator is as a {\it veto}.  This
means that events which might otherwise be used, analyzed or studied
in more detail are rejected because they have a $\chi^2$ value which
is too large.  In general terms, the $\chi^2$ time-frequency
discriminator may be thought of as a method for reducing contributions
from the non-Gaussian tails that characterizes most gravitational-wave
detectors \footnote{If the detector noise were Gaussian, then
histograms of various detection statistics would be power-law or
exponentially distributed. Typically a real detector shows this
behavior in the central part of the distribution, but has a break in
the slope, known as the non-Gaussian tail, at large values.}.
Substantial efforts have been made to characterize these tails in the
TAMA \cite{TamaNonGaussian,TamaNonGaussian2}, Explorer \cite{Gusev},
and Nautilus \cite{Bonifazi} detectors. Other methods for reducing the
effects of these tails have also been proposed and/or used. For example
Creighton \cite{CreightonNonGaussian} proposes a simple analytic model
for non-Gaussian tails and uses it to characterize a network detection
algorithm which is insensitive to this non-Gaussian tail. Other
filtering methods, based on locally-optimal statistics which are less
sensitive or insensitive to non-Gaussian tails have been proposed
\cite{AllenEtAl1,AllenEtAl2} for matched filtering and stochastic
background searches.  Shawhan and Ochsner \cite{ShawhanOchsner} have
developed a heuristic veto method for matched filtering for binary
inspiral, based on counting threshold crossings in a short-time
window.  When tuned for the LIGO S1 data set with a half-second
window, the method is effective, and provides a veto which is
complementary to the $\chi^2$ time-frequency discriminator presented
here.  Some related ideas have also been explored by Guidi
\cite{guidi}.

The principal source that will serve as an example here is the
gravitational radiation back-reaction driven inspiral of pairs of
compact stars, also known as ``binary inspiral''.  If each of the two
stars (typically neutron stars or black holes) has masses smaller than
a few solar masses, then the waveforms can be accurately calculated
over the typical detector bandwidth (30--500~Hz) using post-Newtonian
approximations \cite{lincolnwill, cutleretal, bdiww, biww,%
willwiseman, blanchet:1996, blanchet:1998, blanchet:2001,%
blanchet:2002a, blanchet:2002b}. In this case, the unknown signal
parameters include an overall amplitude scale, the masses and spins of
the two stars, a fiducial reference time (often taken to be the
``coalescence time''), and the initial phase of the orbit.  (The final
unknown parameter, the orbital inclination, is degenerate with these
other parameters, and may therefore be ignored.  The orbital
eccentricity may also be neglected: by the time such systems are
emitting in the detection band, they have radiated away any
eccentricity and the orbit has circularized.)  These signals are
broadband, since the binary system is observed at the very end of its
life when the signal frequency is increasing rapidly as the orbital
period decreases and the stars spiral together.

The paper is structured as follows. Section~\ref{s:conventions}
defines the notational conventions that are used.
Section~\ref{s:matchedfiltering} derives the form of the optimal
matched filter in the simplest case and describes its properties.
Section~\ref{s:discriminator1} defines the $\chi^2$ time-frequency
discriminator and derives its basic properties, also for the simplest
case.  Section~\ref{s:howworks} gives a brief illustrative example of
this statistical test in action, computing and comparing the $\chi^2$
values obtained for a simulated inspiral signal and a spurious noise
event.  Readers interested in acquiring a quick understanding of the
method without seeing the technical details should start with this
Section.

A significant problem in searching for gravitational wave signals is
that the waveform depends upon a number of parameters (for example,
masses) of the source.  For this reason, data must be searched with a
bank of filters designed to cover this parameter space \cite{Owen,OwenSathya}.  Since this
bank is discrete, and the source parameters are continuous, the match
between signal and filter is never exact.  The effects of this
signal/template mismatch on the $\chi^2$ discriminator are
investigated and quantified in Section~\ref{s:mismatch}. One of the
main results of this paper is an absolute upper limit on the expected
value of $\chi^2$ arising from template/signal mismatch.

Up to this point, the signals studied are of the simplest type, which
can be completely described with only two parameters: an overall
amplitude, and an offset/arrival time.  However this is insufficient
for most cases of interest, where the signals are an (a-priori
unknown) linear combination of two different polarizations.
Section~\ref{s:unknownphase} treats this case, deriving two-phase
results analogous to the single-phase results of the previous
Sections.

Section~\ref{s:thresholds} examines suitable thresholds on $\chi^2$
for stationary Gaussian noise, and contrasts these with the heuristic
thresholds used in published analysis of real detector data such as
the LIGO S1 binary inspiral upper limit analysis \cite{ligo_inspiral}.

Section~\ref{s:unequalintervals} examines a variation of the
discriminator based on ``unequal expected SNR'' intervals, and shows
that this discriminator still has most of the properties of the
$\chi^2$ discriminator defined in previous Sections.  .

There are an infinity of possible $\chi^2$-like statistical tests and
discriminators.  Work by Baggio et al. \cite{baggioetal} introduced a
$\chi^2$ test for use with resonant-mass gravitational wave detectors.
In Section~\ref{s:othertests} the $\chi^2$ time-frequency
discriminator of this paper is compared to that test.  While the tests
share some similar features, they have quite different properties and
behavior.

This is followed by a brief Conclusion, which summarizes the main
results and some of the unanswered questions.

Appendix~\ref{s:appendix} contains a short calculation proving that
the time-frequency discriminator defined in this paper has a classical
$\chi^2$ distribution if the detector's noise is Gaussian.
Appendix~\ref{s:max} derives a simple mathematical result used in the
body of the paper,

\section{Conventions}
\label{s:conventions}
The Fourier Transform of a function of time $V(t)$ is denoted by
$\tilde V (f)$ and is given by
\begin{equation}
\label{e:fft1}
{\tilde V}(f) = \int {\rm e}^{- 2 \pi i f t} V(t) dt.
\end{equation}
The inverse Fourier transform is
\begin{equation}
\label{e:fft2}
V(t)= \int {\rm e}^{2 \pi i f t} {\tilde V}(f) df.
\end{equation}
All integrals are from $-\infty$ to $\infty$, unless otherwise
indicated, and ${}^*$ denotes complex conjugate.

The detector output (typically a strain) is denoted by
\begin{equation}
\label{e:detout}
s(t) = n(t) + h(t)
\end{equation}
where $n(t)$ is the (real) strain-equivalent noise produced by
fluctuations within the detector and its environment, and $h(t)$ is a
(real) gravitational waveform of astrophysical origin.

Since the detector's noise $n(t)$ can only be characterized
statistically, one must introduce tools for determining the expected
properties of quantities measured in the presence of this noise.
There are several equivalent ways to do this.  In this paper, we
imagine that $n(t)$ is a random time-series drawn from a large
ensemble of such time series, whose statistical properties are those
of the instrument noise \cite{papoulis}.  If $W$ is some functional that depends upon
$n(t)$, then angle brackets $\langle W \rangle$ denote the average of
$W$ over the ensemble of different $n(t)$.

We assume that $\langle n(t) \rangle $ vanishes, which implies that
$\langle \tilde n(f) \rangle =0$.  We also assume that the statistical
properties of the detector's noise are second-order stationary 
\footnote{Note that we do not need to assume that the process is
stationary.  This would imply that the n-point correlation functions
are time-shift invariant; here we assume only that the two-point
correlation function is time-shift invariant.}  which implies that the
expectation value $\langle n(t) n(t') \rangle$ depends only upon the
time difference $t-t'$. It then follows that in frequency space
\begin{equation}
\label{e:nspec}
\langle \tilde n(f) \tilde n^*(f') \rangle = S_n(f) \delta(f-f'),
\end{equation}
where $\delta(f)$ is the Dirac delta-function.  The {\it two-sided
noise power spectrum} is a real non-negative even function $S_n(f)$
which from (\ref{e:nspec}) can be shown to satisfy
\begin{equation}
\langle n^2(t) \rangle = \int S_n(f) \; df.
\end{equation}
This implies that $ 2 S_n(f)df $ may be interpreted as the expected
squared strain in the frequency band from $f$ to $f+df$, for $f \ge
0$.  (Note that much of the literature on this subject, including
publications of the LIGO Scientific Collaboration, uses a {\it
one-sided} power spectrum $2 S_n(f)$, because this is typically the
quantity measured by standard instrumentation.  Its use here would
complicate many formulae with extraneous factors of two.)

As explained earlier, we are interested in the case of a known
waveform.  A prototypical example is a binary inspiral chirp waveform,
which may be written as \cite{biww}
\begin{equation}
\label{e:waveformtwophase}
h(t) = {D \over d} \bigl( \cos \phi T_c(t-t_0) + \sin \phi T_s(t-t_0)\bigr).
\end{equation}
This waveform depends upon three {\it nuisance parameters}.  These are
the effective distance $d$ to the source, a fiducial time $t_0$ (for
example the coalescence time of the binary pair) and a phase $\phi$
which is determined by the orbital phase of the binary pair and its
orientation relative to the detector.

The templates $T_c$ and $T_s$ are the signal waveforms that would be
produced by a binary inspiral pair optimally oriented with respect to
the detector, at distance $D$, in the two possible polarization states
(corresponding to rotating the detector axes by $45^\circ$).  The
waveform may depend upon additional nuisance parameters; we will
return to this later.

For pedagogic purposes, we first consider the simpler case in which
the phase $\phi$ of the waveform is known a priori, in advance,
\begin{equation}
\label{e:waveform}
h(t) = {D \over d} \; T(t-t_0).
\end{equation}
In this case there are only two nuisance parameters: time of arrival
and effective distance $d$.

The quantity $D$ is the canonical distance at which a source,
optimally-oriented with respect to the detector, would produce the
waveform $T$.  Its value determines the overall normalization scale of
the waveform $T$, since, for a given source type, the quantity $D
T(t)$ is {\it independent} of $D$, and is determined by the laws of
physics, specifically General Relativity.

\section{Matched filtering}
\label{s:matchedfiltering}

A matched filter is the optimal linear filter for detection of a
particular waveform.  Its form can be derived using a number of
different techniques.  Here we use one of the classical signal
analysis methods.

For notational purposes it is useful to introduce a Hermitian inner
product defined by
\begin{equation}
\label{e:hermitian1}
\bigl( A , B \bigr) = \int {A^*(f) B(f)\over S_n(f)} df,
\end{equation}
for any pair of complex functions $A(f)$ and $B(f)$.  The frequency
dependence of $A$ and $B$ will usually be implied and not indicated
explicitly.

A real detector functions only over a finite frequency band, and
acquires data at a finite sample rate.  In this case, the noise power
spectrum $S_n$ may be taken to be infinite outside the bandwidth of
the instrument, effectively restricting the range of integration to
lie between plus and minus the Nyquist frequency $f_N = 1/(2 \Delta
t)$, where $\Delta t$ is the time between successive data samples.

The matched filter is a linear operator that maximizes the ratio of
``signal'' to ``noise''.  We denote the filter by $\tilde Q^*(f)/S_n(f)$ and the
output of the filter by $z$, so
\begin{equation}
\label{e:snr1def}
z \equiv \int { \tilde  Q^*(f) \tilde s(f) \over S_n(f)} \; df =
    (\tilde Q, \tilde s). 
\end{equation}
We require that $z$ be real, which implies that $\tilde Q(f)=\tilde Q^*(-f)$, and
also means that $\tilde Q(f)/S_n(f)$ corresponds to a real function (filter
kernel) in the time domain.

The expected value of $z$ can be found from (\ref{e:waveform}), and is
given by
\begin{equation}
\label{e:eoutput}
\langle z \rangle = {D \over d} \biggl(\tilde Q,
  \tilde T  \;  {\rm e}^{-2 \pi i f t_0} \biggr)
\end{equation}
Here $\tilde T(f)$ denotes the Fourier transform of $T(t)$; the
translation in time by $t_0$ is explicitly encoded in the exponential
term. Note that in the absence of a source ($d \to \infty$) the
expected value of $z$ vanishes since $\langle \tilde n(f) \rangle =0 $
implies that $\langle \tilde s(f) \rangle =0$. 

The expected value of the square of $z$ may be found using
(\ref{e:nspec})
\begin{equation}
\langle z^2 \rangle =
\bigl( \tilde Q , \tilde Q  \bigr) + \bigl( {D \over d}\bigr)^2 \biggl(\tilde Q,
  \tilde T  \;  {\rm e}^{-2 \pi i f t_0} \biggr)^2
\end{equation}
To estimate the error or uncertainty in a measurement of $z$, it is
useful to define
\begin{equation}
\delta z = z - \langle z \rangle.
\end{equation}
The error or uncertainty in a measurement of $z$, due to noise in the
detector, is
\begin{eqnarray}
\label{e:uncertainty}
\nonumber
\sqrt{ \langle (\delta z)^2 \rangle} & = & \sqrt{ \langle \bigl( z - \langle z \rangle \bigr)^2 \rangle} \\
\nonumber
                                     & = & \sqrt{\langle z^2 \rangle - \langle z \rangle^2} \\
                                     & = & \bigl( \tilde Q, \tilde Q \bigr)^{1/2}.
\end{eqnarray}
From these quantities, we can now derive the properties of the optimal
matched filter.

Under the assumptions that we have made about the detector output
(\ref{e:detout}) the optimal choice of matched filter $\tilde Q$ is
the one that maximizes the ratio of the expected filter output $<z>$
given by (\ref{e:eoutput}) to its expected uncertainty
(\ref{e:uncertainty}) due to detector noise.  Hence the optimal choice of $\tilde Q$ maximizes
\begin{equation}
\label{e:rat}
{\langle z \rangle \over 
\sqrt{ \langle (\delta z)^2 \rangle}
 } =  { \bigl( \tilde Q, A \bigr) \over \bigl( \tilde Q , \tilde Q\bigr)^{1/2} },
\end{equation}
where we have {\it defined}
\begin{equation}
A(f) \equiv {D \over d} \tilde T(f)  \; {\rm e}^{- 2 \pi i f t_0}.
\end{equation}
Since the inner product is Hermitian, Schwartz's inequality states that
\begin{equation}
\bigl| \bigl( \tilde Q, A \bigr) \bigr|^2 \le \bigl( A , A \bigr) \bigl( \tilde Q , \tilde Q \bigr).
\end{equation}
The two sides are equal if and only if $\tilde Q$ is proportional to
$A$.  Hence the ratio (\ref{e:rat}) is maximized when $\tilde Q(f)$ is
proportional to $A(f)$.  Thus, the optimal filter is a time-reversed
copy of the template, weighted by the expected noise in the detector
\footnote{ Using (\ref{e:fft1}) one may show that since $\tilde T^*(f)
= \tilde T(-f)$, the quantity $\tilde T^*(f)$ is the Fourier transform
of the time-reversed template $T(-t)$. Hence the quantity $\tilde
Q^*(f)$ that appears in (\ref{e:snr1def}) is a time-reversed image of
the template, weighted by the noise spectrum.}.

The {\it Signal to Noise Ratio} (SNR) is defined by the ratio
of the {\it observed} filter output $z$ to its (expected or observed)
root-mean-square fluctuations
\begin{equation}
{\rm SNR} = {z \over  
\sqrt{ \langle (\delta z)^2 \rangle} } = 
{    (\tilde Q, \tilde s ) \over \sqrt{ \bigl( \tilde Q , \tilde Q\bigr) }},
\end{equation}
and is independent of the normalization of the optimal filter $\tilde Q$.  By
definition, in the absence of a signal $\langle {\rm SNR} \rangle = 0$ and
$\langle ({\rm SNR})^2 \rangle = 1$.

It is convenient to choose the normalization of the optimal filter $\tilde Q$
so that $\bigl( \tilde Q , \tilde Q\bigr) = 1$.  This may be achieved by choosing the
filter $\tilde Q$ to be
\begin{equation}
\label{e:firstnorm}
\tilde Q(f)  = 
\bigl( \tilde T  , \tilde T \bigr)^{-1/2} \;
\tilde T(f) \; {\rm e}^{-2 \pi i f t_0}.
\end{equation}
With this normalization choice, the filter output $z$ is equal to the
SNR.  Henceforth we will use $z$ to denote both of these quantities.


While the optimal filter $\tilde Q$ is explicitly independent of the
normalization scale of the template $T$, we showed earlier that the
scales of $D$ and $T$ could be freely adjusted provided that their
product $D T$ was held fixed.  For the purposes of interpreting the
SNR $z$, it is convenient to set the distance scale $D$ so that
\begin{equation}
\label{e:secondnorm}
\bigl( \tilde T , \tilde T \bigr) = 1.
\end{equation}
With this choice of normalization, the expected value of the SNR is
\begin{equation}
\langle z \rangle = {D \over d}
       \bigl( \tilde T , \tilde T \bigr)^{1/2} = {D \over d}.
\end{equation}
This choice of normalization is thus equivalent to choosing the
distance $D$ at which the template is defined to be the distance at
which an optimally-oriented source would have an expected SNR of
unity: $\langle z \rangle = 1$.

Since the expected value of $z$ is proportional to the inverse
distance, one may use the actual measured value of $z$ to estimate the
distance.  Since the actual measured value of $z$ is affected by
instrument noise, this estimator has some average error.  One can
easily estimate the error, since with our normalization choices
\begin{eqnarray}
\nonumber
\langle z^2 \rangle & =  & 1 + \biggl( {D \over d } \biggr)^2, \text{ and hence} \\
\langle ( \delta z)^2 \rangle & = & 1.
\end{eqnarray}
This means that the expected fractional error in estimating the
inverse distance to the source is
\begin{equation}
{ \langle ( \delta z)^2 \rangle^{1/2} \over \langle z \rangle } = {1 \over \langle z \rangle } = {d \over D}
\end{equation}
Thus, a measured SNR of $z=10$ implies a fractional accuracy in
distance determination of about 10\%.

Up to this point, we have been assuming that the fiducial coalescence
time $t_0$ is known.  In practice, one searches a data stream for
statistically significant values of $z$, for all possible choices of
$t_0$.  As a function of $t_0$ the SNR is
\begin{equation}
z(t_0) = \int { \tilde s(f) \tilde T^* \; {\rm e}^{2 \pi i f t_0}
 \over S_n(f)} \; df.
\end{equation}
Because this is just an inverse Fourier transform, it is both
practical and simple to compute this quantity from a data stream
$s(t)$. For example the Fast Fourier Transform (FFT) algorithm allows
the r.h.s. to be computed in order $N \ln N$ operations, where $N$ is
the number of data samples of $h$ in the time or frequency domain.

\section{The $\chi^2$ discriminator test}
\label{s:discriminator1}

In the previous section, we assumed only that the detector noise was
second-order stationary.  It is quite common in such studies to also
assume that the noise is Gaussian.  One may then show that the
probability of the SNR exceeding some threshold falls exponentially
with increasing threshold, and so large values of the SNR have low
probability of being due to noise in the detector, and thus are a good
indication that a real source is present.

Unfortunately, experience has shown that the noise in broadband
gravitational wave detectors is far from Gaussian. Typically it has a
Gaussian or Gaussian-like component (arising from electrical, thermal
and shot noise) but there is another ``glitchy'' component that could
be described as Poisson-like impulse noise.  There are many sources of
this noise, including marginally stable servo systems and
environmental anomalies.  The effects of this noise on the filtering
process described above can be quite dramatic.  Whereas the matched
filter is designed to give a large response when the signal waveform
matches the template, it also can give a large response when the
instrumental noise has a large glitch.  Although the waveform of this
glitch looks nothing like the template, it nevertheless drives the
filter output to a large value.

The statistical test described here provides a way to determine if the
output of the filter is consistent with what might be expected from a
signal that matched the template.  To be effective, both the signal
and the detector must be broadband.

One way to understand how this test works is to imagine that instead
of a single broadband detector, one is given $p$ data streams from $p$
different, independent narrow-band detectors, each operating in a
different frequency band.  For each detector, one can construct an
optimal filter for the signal, and then one can ask if the results are
consistent, for example, if the $p$ (potentially different) fiducial
times $t_0$ which maximize the output of each of the $p$ independent
detectors are consistent with a single value.

Begin by assuming that, using matched filtering as previously
described, we have identified a time of arrival $t_0$ and inverse
distance $D/d$. The goal is to construct a statistic which indicates
if the filter output is consistent with this signal.

We will do this by investigating the way in which $z(t_0)$ gets its
contribution from different ranges of frequencies.  To do this, we
partition the frequency range $f \in [0,\infty) $ into a set of $p$
distinct subintervals $\Delta f_1,\cdots,\Delta f_p$ whose union is
$[0,\infty)$. The frequency intervals:
\begin{eqnarray}
\label{e:freqintervals}
\nonumber
\Delta f_1 &=& \{ f \; | \; 0 \le f < f_1 \} \\*
\nonumber
\Delta f_2 &=& \{ f \; | \; f_1 \le f < f_2 \} \\*
\nonumber
\cdots \\*
\nonumber
\Delta f_{p-1} &=& \{ f \; | \; f_{p-2} \le f < f_{p-1} \} \\*
\Delta f_p &=& \{ f \; | \; f_{p-1} \le  f < \infty \},
\end{eqnarray}
will be defined by the condition that the {\it expected signal
contributions in each frequency band from a chirp are equal}. (Note
that one may also pick intervals which do not satisfy this condition.
In Section~\ref{s:unequalintervals} we show that when suitably
defined, the resulting statistic still has a classical $\chi^2$
distribution for the case of Gaussian detector noise.)

To define the frequency bands, it is helpful to introduce a set of $p$
Hermitian inner products (for $j=1,\cdots,p$) defined in analogy to
(\ref{e:hermitian1}) by
\begin{equation}
\label{e:hermitian2}
\bigl( A , B \bigr)_j = \int_{-\Delta f_j \cup \Delta f_j} {A^*(f) B(f)\over S_n(f)} \; df.
\end{equation}
In each of these integrals, the range of integration is over both the
positive and negative frequencies.  As discussed following
(\ref{e:hermitian1}), since $S_n(f)$ may be taken as infinite for
$|f|$ greater than the Nyquist frequency $f_N$, the effective upper
limit of the final frequency interval $\Delta f_p$ is $f_N$ rather
than $\infty$.

\begin{center}
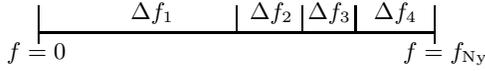
\begin{figure}
\begin{center}
\begin{picture}(150,20)(0,0)
\put(0,10){\line(1,0){150}} \put(0,6){\line(0,1){14}}
\put(75,10){\line(0,1){10}} \put(35,15){$\Delta f_1$}
\put(100,10){\line(0,1){10}} \put(80,15){$\Delta f_2$}
\put(120,10){\line(0,1){10}} \put(102,15){$\Delta f_3$}
\put(150,6){\line(0,1){14}} \put(127,15){$\Delta f_4$}
\put(-12,0){$f=0$} \put(138,0){$f=f_{\rm Ny}$}
\end{picture}
\end{center}
\caption{
\label{f:fintervals}
A typical set of frequency intervals $\Delta f_j$ for the case $p=4$.
These intervals are narrowest where the detector is the most
sensitive, and broadest where it is least sensitive.}
\end{figure}
\end{center}

Since the frequency intervals don't overlap, but cover all frequency
values, the sum of these inner products
\begin{equation}
\bigl( A , B \bigr) = \sum_{j=1}^p \bigl( A , B \bigr)_j
\end{equation}
yields the inner product (\ref{e:hermitian1}) defined earlier.  The
$p$ distinct frequency bands are uniquely determined by the condition
that
\begin{equation}
\label{e:howtosetintervals}
\text{choose } \Delta f_j \text{ so that }
\bigl( \tilde T , \tilde T)_j = {1 \over p}.
\end{equation}
A typical set of frequency intervals in shown in Figure \ref{f:fintervals}.

For given instrumental noise $S_n(f)$ the frequency intervals $\Delta
f_j$ depend upon the template $T$.  However it may be the case that
many templates actually share the {\it same} frequency intervals
$\Delta f_j$.  A good example of this is the set of stationary-phase
post-Newtonian templates, where the amplitude is calculated in the
first post-Newtonian approximation, and the phase is calculated to
higher order \cite{stationaryphase1,stationaryphase2}.
For these templates, the frequency intervals are determined by
\begin{eqnarray}
\label{e:statphase}
(\tilde T, \tilde T)_j & = & {1 \over p} (\tilde T, \tilde T) \\
\nonumber
            \int_{\Delta f_j} {f^{-7/3} \over S_n(f) } \; df & = & 
{1 \over p} \int_0^\infty {f^{-7/3} \over S_n(f) } \; df
\end{eqnarray}
provided that $m_1$ and $m_2$ lie in a range for which the stationary
phase approximation holds within the detector band \footnote{ For
binary systems, the stationary-phase approximation is very accurate at
low frequencies.  However the stationary-phase approximation breaks
down somewhat below twice the orbital frequency $\Omega_{\rm lsco}$ of
the last stable circular orbit.  Thus if $2 \Omega_{\rm lsco}$ lies
within or below the highest frequencies for which the detector is
sensitive, then the stationary phase approximation and equation
(\ref{e:statphase}) do not hold.  }.  For this family of templates,
{\it all} the templates share the {\it same} bands $\Delta f_j$.

The SNR (\ref{e:snr1def}) is an integral over all frequencies, and can
be written as a sum of contributions from each of the $p$ different
bands,
\begin{equation}
z = \sum_{j=1}^p z_j \text{ with } z_j \equiv (\tilde Q, \tilde s )_j.
\end{equation}

The expected values of $z_j$ and its square are computed using the
same techniques as before, and give
\begin{eqnarray}
\label{e:expect1}
\nonumber
\langle z_j   \rangle & = & {1 \over p} {D \over d}  \text{, and} \\
\langle z_j^2 \rangle & = & {1 \over p} + {1 \over p^2 } \bigl( {D \over d}\bigr)^2
\end{eqnarray}
In the absence of a signal (take $d \to \infty$) one finds
\begin{equation}
\label{e:expect2}
\langle z_j   \rangle  = 0 \text{ and } \langle z_j^2 \rangle = {1 \over p}.
\end{equation}
This suggests an obvious statistical test to see if the signal is
consistent with the model.

Consider the $p$ quantities defined by
\begin{equation}
\Delta z_j \equiv z_j - { z \over p}.
\end{equation}
These are the differences between the SNR in the band $\Delta f_j$,
and the SNR that would be anticipated \footnote{\label{n:anticipated}
The word ``anticipated'' is used, rather than ``expected'', because we
do not experimentally or observationally have access to the expected
value of the SNR $\langle z \rangle$.  In other words, the quantities
$\Delta z_j = z_j - z/p $ are {\it not} the same as $\delta z_j = z_j
- \langle z_j \rangle = z_j - \langle z \rangle/p $.  For any given
observation or data set, the quantities $\delta z_j$ do {\it not} sum
to zero (although their expectation value does).  They have expected
square values $\langle (\delta z_j )^2 \rangle = 1/p$.} in that band,
based on the total {\it measured} SNR in all bands.  By definition,
these differences sum to zero
\begin{equation}
\sum_{j=1}^n \Delta z_j = 0
\end{equation}
and their individual expectation values vanish
\begin{equation}
\langle \Delta z_j \rangle =0.
\end{equation}
To calculate the expectation values of their squares, first note that
the quantity $\langle z_j z \rangle$ must, by symmetry, be
$j$-independent \footnote{The assumption that the noise is second
order stationary and the fact that $\Delta f_j$ and $\Delta f_k$ do
not overlap for $j \ne k$ implies that $\langle z_j z_k \rangle$
vanishes for $j \ne k$.}.  Since the sum over $j$ of $\langle z_j z
\rangle$ yields $\langle z^2 \rangle$, one must have
\begin{equation}
\langle z_j z \rangle = {\langle z^2 \rangle \over p} = {1 \over p}
\biggl[ 1 + \biggl( {D \over d } \biggr)^2 \biggr].
\end{equation}
Thus the expectation value of the square of $\Delta z_j$ is
\begin{eqnarray}
\label{e:expectdelta2}
\nonumber
\langle (\Delta z_j)^2 \rangle & = & \langle \bigl(z_j - {z \over p}\bigr)^2 \rangle \\
\nonumber
 & = & \langle z_j^2 \rangle + {\langle z^2 \rangle \over p^2}  - {2 \langle z_j z \rangle \over p}  \\ 
& = & {1 \over p} \biggl( 1 - {1 \over p} \biggr)
\end{eqnarray}
Notice that these quantities do not depend upon $d$.  In fact {\it
these quantities,and their second-order statistical properties, are
independent of whether or not a signal is present}.  This motivates
the definition of a discrimination statistic.

We define the $\chi^2$ time-frequency discriminator statistic by
\begin{equation}
\label{e:definechisquare}
\chi^2 = \chi^2 (z_1, \cdots, z_p) = p \sum_{j=1}^p (\Delta z_j)^2.
\end{equation}
This choice of statistic is one of the main results of the paper: in
what follows we will study its properties in detail.

It follows immediately from (\ref{e:expectdelta2}) that the expected
value of $\chi^2$ is
\begin{equation}
\langle \chi^2 \rangle = p -1
\end{equation}
Up to this point, the only assumption we have made is that the noise
in the instrument is second order stationary, specifically that
$\langle n(t) n(t') \rangle$ depends only upon $t-t'$.  To further
analyze the properties of this statistic, additional assumptions are
needed.

In the design of signal processing algorithms, it is common to analyze
the performance of the method in the case where the instrument noise
is both stationary and Gaussian.  In this case,
Appendix~\ref{s:appendix} shows that probability distribution function
of $\chi^2$ is a a classical $\chi^2$-distribution with $p-1$ degrees
of freedom.  The (cumulative) probability that $\chi^2 < \chi_0^2$ is
\begin{eqnarray}
P_{\chi^2 < \chi_0^2} & = & \int_0^{\chi_0^2/2} { u^{{p \over 2}-{3
\over 2}} {\rm e}^{-u} \over \Gamma({p \over 2}-{1 \over
2})} \; du \\ & = & {\gamma({p \over 2}-{1
\over 2}, { \chi_0^2 \over 2}) \over \Gamma({p \over 2}-{1 \over 2})}
\end{eqnarray}
where $\gamma$ is the incomplete gamma function.  In this case, where
the noise is assumed to be stationary and Gaussian, the expected
distribution of $\chi^2$ values is quite narrow. One has
\begin{equation}
\langle (\chi^2)^2 \rangle = p^2 - 1
\end{equation}
which implies that the ``width'' of the $\chi^2$ distribution is
\begin{equation}
\bigl( \langle (\chi^2)^2 \rangle - \langle \chi^2 \rangle^2 \bigr)^{1/2} = \sqrt{2 (p-1)}
\end{equation}
Thus, if the noise were stationary and Gaussian, we would expect to
find $\chi^2$ values in the range $[p-1-\sqrt{2 (p-1)}, p-1+\sqrt{2
(p-1)}]$.  Since the fractional width of this range decrease with
increasing $p$, one might expect that large values of $p$ are
desirable, since they appear to give high discriminating power.

Practice and experience have shown that large values of $p$ do not, in
fact, work very well \cite{Babak}. Partly this is because the detector
noise is neither stationary nor Gaussian, and partly this is because
the signal is not a perfect match to the template. Large values of $p$
tend to spread non-stationary glitch noise over many frequency bands,
diluting its effect on $\chi^2$. This is difficult or impossible to
model analytically, and can best be understood (as in \cite{Babak}) by
Monte-Carlo studies of simulated signals added into real detector
noise.  However the effects of a signal-template mismatch can be
studied analytically; this is done in Section~\ref{s:mismatch}.

\section{How does the $\chi^2$ test work?}
\label{s:howworks}
\begin{figure}
\begin{center}
\epsfig{file=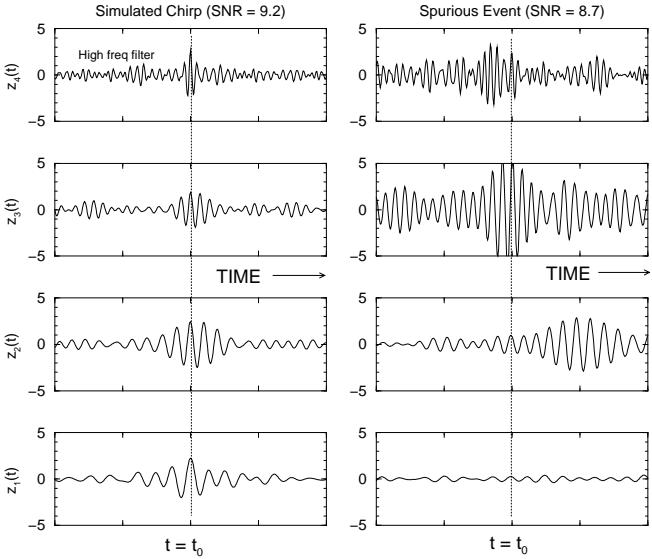,width=2.9in,angle=-90}
\caption{
\label{f:timefreqplot}
The output of $p=4$ single-phase filters for a simulated chirp signal
added into a stream of detector noise (left set of figures) and a
transient burst present in detector noise (right set of figures). For
the simulated chirp, the filters in the different frequency bands all
peak at the same time offset $t_0$: the time offset which maximizes
the SNR.  At this instant in time, all of the $z_j$ are about the same
value.  However when the filter was triggered by the transient burst,
the filters in the different frequency bands peak at different
times. At time $t_0$ they have very different values (some large, some
small, and so on).  }
\end{center}
\end{figure}

The $\chi^2$ test was invented based on experience filtering data from
the LIGO 40m prototype instrument \cite{grasp_chi2,40inspiral}.  It
was observed that a binary inspiral filter bank registered many events
that (when converted to audio) did not sound like inspiral signals.
In particular, the low frequency component of the signal did not
arrive first, followed by the midrange and high frequency components.
The $\chi^2$ test first arose from considering a set of matched
filters in different bands, and testing to see if the filter outputs
all peaked at the correct time.  The signal $z_1$ was constructed from
the lowest frequency band, $z_2$ from the next frequency band, and so
on.  This is illustrated in Figure~\ref{f:timefreqplot} for a
single-phase test with $p=4$ bands.

It is instructive to compare the values of the filter outputs
(single-phase test) for the two cases shown in Figure
\ref{f:timefreqplot}.  For the simulated chirp, the signal-to-noise
ratio was $z=9.2$ and the signal values in the different bands were
\begin{eqnarray}
\nonumber
z_1 & = & 2.25\\
\nonumber
z_2 & = & 2.44\\
\nonumber
z_3 & = & 1.87\\
\nonumber
z_4 & = & 2.64\\
z & = & z_1 + z_2 + z_3 + z_4 = 9.2 \\
\nonumber
\chi^2 & = & 4 \sum_{j=1}^4 (z_j - z/4)^2 = 1.296\\
\nonumber
P_{\chi^2 \ge 1.296} & = & 1 - { \gamma(3/2, 0.648 ) \over \Gamma(3/2)} = 73\%.
\end{eqnarray}
This is quite consistent with the value of $\chi^2$ that would be
expected for a chirp signal in additive Gaussian noise.

For the spurious noise event shown in Figure \ref{f:timefreqplot} the
SNR $z=8.97$ was quite similar but the value of $\chi^2$ is very
different:
\begin{eqnarray}
\nonumber
z_1 & = & 0.23\\
\nonumber
z_2 & = & 0.84\\
\nonumber
z_3 & = & 5.57\\
\nonumber
z_4 & = & 2.33\\
z & = & z_1 + z_2 + z_3 + z_4 = 8.97\\
\nonumber
\chi^2 & = & 4 \sum_{j=1}^4 (z_j - z/4)^2 = 68.4\\
\nonumber
P_{\chi^2 \ge 68.4} & = & 1- { \gamma (3/2, 34.2) \over \Gamma(3/2)} =9.4 \times 10^{-15}.
\end{eqnarray}
The probability that this value of $\chi^2$ would be obtained for a
chirp signal in additive Gaussian noise is {\it extremely} small.

\section{Effect of a signal/template mismatch on $\chi^2$}
\label{s:mismatch}

In the previous two Sections, we analyzed an optimal filter and
constructed a $\chi^2$ statistic for the case where the signal
waveform was known exactly.  In practice, this is not possible.
Typically, signal waveforms come from a family characterized by a set
of continuous parameters, such as masses and spins.  Thus, in
practice, to search for signals one uses a discrete set of templates,
called a {\it template bank} \cite{Owen,OwenSathya}.  Such banks can
contain anywhere from dozens to hundreds of thousands of templates.
Since each template in the bank is defined by a point in parameter
space, the template bank may be thought of as a grid, or mesh, in
parameter space.  Typically, this grid is laid out to ensure that any
signal from the continuous family of waveforms is ``near'' some point
in the grid.  In this section, we analyze the case where the signal
waveform is ``close'' to the template waveform, but not a perfect
match.

We begin by assuming that the signal is perfectly described by a
template $T'$, so that the detector's output is
\begin{equation}
 s(t) = n(t) + {D' \over d'} \; T'(t).
\end{equation}
Adopting the same conventions as before, we assume that $D'$ is chosen
so that $T'$ obeys $(\tilde T', \tilde T') = 1$. For simplicity, and
without loss of generality, we take $t_0=0$.  Assume that this signal
is ``close'' to that of the template $T$, and hence that the signal is
detected in that template.  The SNR is
\begin{equation}
z = (\tilde Q, \tilde s) = (\tilde T, \tilde n) + {D' \over d'} (\tilde T, \tilde T').
\end{equation}
Using Schwartz's inequality, the inner product between the two
templates must lie in the range $[-1,1]$.
\begin{eqnarray}
\nonumber \bigl( \tilde T, \tilde T')^2 & \le & \bigl( \tilde T,
 \tilde T) \bigl( \tilde T', \tilde T') \\
& \le & 1.
\end{eqnarray}
One may think of the two templates as unit vectors separated by an
angle $\theta$ and write this in the form
\begin{equation}
\label{e:totalnorm}
\nonumber \bigl( \tilde T, \tilde T') =  \cos \theta, \text{  for  } \theta \in [0, \pi].
\end{equation}
This inner product is often called the {\it fitting factor}. The
expected value of the SNR
\begin{equation}
\langle z \rangle = {D' \over d'} \cos \theta,
\end{equation}
is reduced by a factor of the fitting factor compared with the
expected SNR $D'/d'$ that would be obtained if the template bank
contained the perfectly matching template $T'$.

The fractional difference between this ``ideal case'' expected SNR and
the expected SNR in the mismatched template is called the {\it
template mismatch} $\epsilon$
\begin{equation}
\cos \theta = 1 - \epsilon.
\end{equation}
The value of $\epsilon$ must lie in the range $\epsilon \in [0,2]$,
and may be restricted to the range $\epsilon \in [0,1]$ by changing
the sign of $T'$ if needed.  Hence, without loss of generality we will
assume that $0 \le \cos \theta \le 1$ and that $0 \le \epsilon \le 1$.
The case of most interest is when $\epsilon \ll 1$.  Typically
template banks are set up so that the worst-case mismatch corresponds
to a a loss of event rate (for a uniform source distribution) of 10\%.
Since the volume inside a sphere of radius $r$ grows proportional to
$r^3$ and the SNR is inversely proportional to distance, this
corresponds to a typical worst-case template mismatch of $\epsilon=
0.033 = 3.3\%$.

Following the same procedures as in
Section~\ref{s:matchedfiltering} one can find the expected SNR
squared, which is
\begin{equation}
\langle z^2 \rangle = 1 + \bigl( {D' \over d'} \bigr)^2 \cos^2 \theta .
\end{equation}
Thus, the first- and second-order statistics of $z$ are
indistinguishable from those that would be produced by a signal from a
perfectly matched template $T$ with SNR $D' \cos \theta/d'$.

To analyze the effects of the signal/template mismatch on the $\chi^2$
statistic is slightly more involved.  We begin by considering the way
in which the templates overlap in each individual frequency band.
Define a set of $p$ real constants $\lambda_1, \cdots, \lambda_p$ by
\begin{equation}
\bigl( \tilde T, \tilde T' \bigr)_j = \lambda_j \cos \theta.
\end{equation}
It follows from (\ref{e:totalnorm}) that these constants sum to unity,
\begin{equation}
\sum_{j=1}^p \lambda_j = 1.
\end{equation}
The average value of the $\lambda_j$ is $1/p$.  The deviation away
from this value is a measure of how close together (or far apart) the
templates $T$ and $T'$ are in the frequency band $\Delta f_j$.

The goal is to understand how the $\chi^2$ statistic is affected by
signal/template mismatch.  To determine this, we first express the SNR
in the $j$'th band as
\begin{eqnarray}
z_j & = & (\tilde Q, \tilde s)_j = (\tilde T, \tilde n)_j + {D' \over
d'} (\tilde T, \tilde T')_j \\ & = & (\tilde T, \tilde n)_j + {D'
\over d'} \lambda_j \cos \theta.
\end{eqnarray}
Using calculations identical to Section~\ref{s:discriminator1} the
expected value of the SNR and its square in the $j$'th band are
\begin{eqnarray}
\langle z_j \rangle & = & {D' \over d'} \lambda_j \cos \theta, \text{
and }\\
\langle z^2_j \rangle & = & {1 \over p} + \biggl( {D' \over d'}
\biggr)^2 \lambda^2_j \cos^2 \theta.
\end{eqnarray}
As before, we define $\Delta z_j \equiv z_j - z/p$, giving
\begin{equation}
\Delta z_j = (\tilde T, \tilde n)_j - {1 \over p} (\tilde T, \tilde n) + 
{D' \over d'} (\lambda_j - {1 \over p}) \cos \theta.
\end{equation}
The first difference between this analysis and the one for matching
templates is that in the mismatched case, the expectation value of
$\Delta z_j$ does not vanish:
\begin{equation}
\label{e:nonvanishing}
\langle \Delta z_j \rangle = 
{D' \over d'} \bigl(\lambda_j - {1 \over p} \bigr) \cos \theta.
\end{equation}
As before, we can use the assumption that the detector noise is
second-order stationary to calculate
\begin{eqnarray}
\nonumber
\langle z_j z \rangle & = & \biggl\langle \bigl[ (\tilde T, \tilde n)_j + {D' \over d'} \lambda_j \cos \theta \bigr] \\ 
\nonumber
& & \times \bigl[  (\tilde T, \tilde n) + {D' \over d'} \cos \theta \bigr] \biggr\rangle \\
\nonumber
& = & (\tilde T, \tilde T)_j + \biggl( {D' \over d'} \biggr)^2 \lambda_j \cos^2 \theta \\
& = & {1 \over p} + \biggl( {D' \over d'} \biggr)^2 \lambda_j \cos^2 \theta.
\end{eqnarray}
Using these results, it is straightforward to work out the expectation
value of $(\Delta z_j)^2$:
\begin{eqnarray}
\langle (\Delta z_j)^2 \rangle
& = & \langle z_j^2 \rangle + {\langle z^2 \rangle \over p^2} -{2 \langle z_j z \rangle \over p} \\
\nonumber
& = & {1 \over p} \bigl(1-{1 \over p} \bigr) + \biggl( { D' \over d'}\biggr)^2
 \bigl( \lambda_j - {1 \over p} \bigr)^2 \cos^2 \theta .
\end{eqnarray}
The expectation value of the $\chi^2$ discriminator statistic
(\ref{e:definechisquare}) is therefore
\begin{eqnarray}
\nonumber
\langle \chi^2 \rangle
  & = & p - 1 + \biggl( { D' \over d'}\biggr)^2  \cos^2
         \theta \sum_{j=1}^p p \bigl(\lambda_j - {1 \over p} \bigr)^2 \\
  & = & p - 1 + \langle z \rangle^2 \sum_{j=1}^p p \bigl(\lambda_j - {1 \over p} \bigr)^2.
\end{eqnarray}
This is in sharp contrast to the case where the signal matched the
template perfectly.  In that case, the expectation value of $\chi^2$
was independent of the signal strength. Here, when the signal and
template do not match perfectly, the expected value of $\chi^2$ depends
quadratically on the expected SNR $\langle z \rangle$ of the signal.

The dependence of the discriminator $\langle \chi^2 \rangle$ on the
square of the expected SNR $\langle z \rangle^2 $ has a coefficient
\begin{equation}
\label{e:kappa}
\kappa  = p \sum_{j=1}^p \bigl(\lambda_j - {1 \over p} \bigr)^2 = -1 + p \sum_{j=1}^p \lambda_j^2.
\end{equation}
The quantity $\kappa$ is manifestly non-negative; we now obtain an
absolute upper bound on its value.  Schwartz's inequality implies that
\begin{eqnarray}
\label{e:schwartzbound}
\nonumber
 \bigl( \tilde T, \tilde T' \bigr)_j^2 & \le &   \bigl( \tilde T, \tilde T \bigr)_j  \bigl( \tilde T', \tilde T' \bigr)_j \\
\nonumber
\lambda^2_j \cos^2 \theta & \le &  {1 \over p }\bigl( \tilde T', \tilde T' \bigr)_j \\
\nonumber
& \Rightarrow &  \\
\lambda_j^2 & \le & {1 \over p \cos^2 \theta } \bigl( \tilde T', \tilde T' \bigr)_j.
\end{eqnarray}
Summing both sides over $j$ one obtains
\begin{equation}
\sum_{j=1}^p \lambda_j^2 \; \le \; {1 \over p \cos^2 \theta }.
\end{equation}
Combining this with the definition (\ref{e:kappa}) of $\kappa$, one
obtains the bound
\begin{equation}
\label{e:kappalimit1}
0 \; \le \; \kappa \; \le \; {1 \over \cos^2 \theta} -1.
\end{equation}
Note that this relationship does not assume {\it any} relationship
between the signal waveforms $T$ and $T'$.

For the case of most interest (small template mismatch $\epsilon \ll
1$) this yields the bound
\begin{equation}
\label{e:kappalimit1e}
0 \; \le \; \kappa \; \le \; 2 \epsilon
\end{equation}
and hence one of the main results of this paper
 \footnote{A relationship
of this form was obtained independently by Jolien Creighton.}
\begin{equation}
\label{e:mainresult1}
\langle \chi^2 \rangle = p - 1 + \kappa \langle z \rangle^2 \text{ with
} 0 \; \le \; \kappa \; \le \; 2 \epsilon.
\end{equation}
This relationship {\it only} assumes that the fitting factor between
the signal waveforms $T$ and $T'$ is close to one.

We can obtain a different and tighter bound on $\kappa$ if we assume
that the frequency bands defined by (\ref{e:freqintervals}) and
(\ref{e:howtosetintervals}) are the {\it same} for the waveforms $T$
and $T'$.  For example, as discussed earlier in the context of
equation ({\ref{e:statphase}), this is true for binary inspiral
waveforms in the stationary-phase approximation.  In this case, since
$(\tilde T' , \tilde T')_j = 1/p$, equation (\ref{e:schwartzbound})
implies that
\begin{equation}
\lambda_j^2 \cos^2 \theta \le {1 \over p^2}.
\end{equation}
Thus one has
\begin{equation}
{-1 \over p \cos \theta} \le \lambda_j \le {1 \over p \cos \theta}.
\end{equation}
It is convenient to define $\omega_j \equiv 1/p - \lambda_j$.  In
terms of these quantities
\begin{equation}
\kappa = p \sum_{j=1}^p \omega_j^2.
\end{equation}
The values of $\omega_j$ are constrained by two relations:
\begin{eqnarray}
 \sum_{j=1}^p & \omega_j & = 0, \text{  and} \\
  {1 \over p}(1 - { 1  \over \cos \theta} ) \le & \omega_j & \le
   {1 \over p}(1 + {  1 \over \cos \theta} ).
\end{eqnarray}
If we assume that $p>2$ and
\begin{equation}
\label{e:boundary1}
\epsilon = 1 - \cos \theta \le {2 \over p}
\end{equation}
then the maximum of $\kappa$ is obtained when
\begin{eqnarray}
\nonumber
\omega_1 & = & - {p-1 \over p}\biggl( 1 - { 1 \over \cos \theta} \biggr), \text{  and}\\
\label{e:convexresult}
\omega_2 & = & \cdots = \omega_p = 
{1 \over p}\biggl( 1 - { 1 \over \cos \theta}  \biggr).
\end{eqnarray}
The upper bound on $\kappa$ is
\begin{equation}
\label{e:kappalimit2}
\kappa \le (p-1) \biggl( { 1 \over \cos \theta} -1 \biggr)^2 \sim (p-1) \epsilon^2,
\end{equation}
where in the final part of the relation we assume as before that the
mismatch $\epsilon \ll 1$.  This is one of the other main results of
the paper.  In the case where the bands $\Delta f_j$ used to calculate
$\chi^2$ are the {\it same} for the template and the actual signal, $p
\le 2/\epsilon $, and $\epsilon$ is small, one has
\begin{equation}
\label{e:mainresult2}
\langle \chi^2 \rangle = p - 1 + \kappa \langle z \rangle^2 \text{ with
} 0 \; \le \; \kappa \; \le \; (p-1) \epsilon^2.
\end{equation}
This result is beautifully consistent with the previous limit on
$\langle \chi^2 \rangle$. At the boundary of validity
(\ref{e:boundary1}) of (\ref{e:kappalimit2}), one has
\begin{equation}
\kappa = (p-1)  \biggl( { 1 \over \cos \theta} -1 \biggr)^2 = 
{ 1 \over \cos^2 \theta } -1 ~ \sim 2 \epsilon
\end{equation}
which agrees {\it exactly} with the previous limit
(\ref{e:kappalimit1}).  The limits that apply are summarized in
Figure~\ref{f:limits}.  Note that when the two different limits
(\ref{e:kappalimit1}) and (\ref{e:kappalimit2}) are expressed
approximately to lowest order in terms of $\epsilon$, they appear to
differ slightly at the boundary $p=2/\epsilon$.  In fact they agree
exactly: the approximate expressions differ at higher order in
$\epsilon$.

\begin{figure}
\begin{center}
\epsfig{file=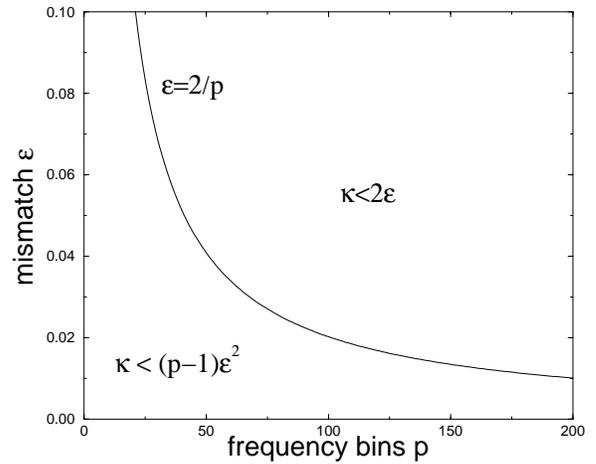,angle=-90,width=3in}
\caption{ \label{f:limits} The expected value of $\chi^2$ for a signal
that is not matched to the template satisfies $\langle \chi^2 \rangle
= p-1 + \kappa \langle z \rangle^2$ where $\langle z \rangle$ is the
expected SNR.  If signal and template share the same set of frequency
bands, then below the curve $\epsilon=2/p$ one has $\kappa < (p-1)
\epsilon^2$.  Above the curve $\kappa < 2 \epsilon $.  If signal and
template do {\it not} share the same frequency bands, then $\kappa < 2
\epsilon $ is the only limit that applies.}
\end{center}
\end{figure}

In Appendix~\ref{s:appendix} show that if the detector noise is
Gaussian, then the $\chi^2$ statistic (computing with a perfectly
matching template) has a classical $\chi^2$-distribution with $p-1$
degrees of freedom.  If the template does {\it not} match perfectly,
the distribution becomes a non-central $\chi^2$ distribution, with a
non-centrality parameter determined by the r.m.s. value of
(\ref{e:nonvanishing}), which is $\kappa \langle z \rangle^2$.  This
is discussed in more detail at the end of
Section~\ref{s:unknownphase}. It follows from the fact that the
variance of the terms that appear in the definition of $\chi^2$,
\begin{equation}
\langle (\Delta z_j )^2 \rangle - \langle \Delta z_j \rangle^2 = {1 \over p} \bigl(1-{1 \over p} \bigr)
\end{equation}
are independent of the signal amplitude.

\section{Signal of unknown phase}
\label{s:unknownphase}
As mentioned earlier, the signal from an inspiraling binary pair is a
linear combination of two possible gravitational waveforms with an
a-priori unknown phase $\phi$.  Here, we repeat the analysis done in
the previous three sections, for this particular case of interest.

As in (\ref{e:waveformtwophase}), the detector output is assumed to be
of the form
\[
\nonumber
s(t) = n(t) + {D \over d} \bigl( \cos \phi T_c(t-t_0) + \sin \phi T_s(t-t_0)\bigr),
\]
with $n(t)$ a random time-series drawn from a distribution appropriate
to the detector noise, and both $\phi$ and $d$ unknown.  We assume (as
is the case for binary inspiral) that the templates are orthonormal so
that \begin{eqnarray}
\bigl( \tilde T_c,  \tilde T_c \bigr) & = & \bigl( \tilde T_s,  \tilde T_s \bigr) = 1, \text{ and} \\
\bigl( \tilde T_c,  \tilde T_s \bigr) & = & 0.
\end{eqnarray}
Note that in the stationary phase approximation, $T_c$ and $T_s$ are
exactly orthogonal.  Were they not, a Gram-Schmidt procedure could
be used to construct an orthonormal pair spanning the same space of
signals \footnote{One can construct an orthonormal $\tilde T_s$
starting with time domain waveforms by multiplying $\tilde T_c$ by $i$
for positive frequencies and $-i$ for negative frequencies.  However
this is {\it not} exactly the Fourier transform of the time-domain
$T_s$.}.

There are several (easy) ways to efficiently search for the unknown
phase $\phi$.  In substance, all of these methods consist of filtering
{\it separately} with the two templates $T_c$ and $T_s$, and then
combining the two filtered data streams.  For our purposes a nice way
to do this is to combine these separate (real) filter outputs into a
single complex signal.  Thus, we use the optimal filter
\begin{equation}
\tilde Q = \bigl( \tilde T_c + i \tilde T_s \bigr) {\rm e}^{-2\pi i f t_0}.
\end{equation}
Note that with this normalization the optimal filter is normalized so
that $\bigl( \tilde Q, \tilde Q \bigr)=2$.

The output of the filter is complex and is
\begin{eqnarray}
\nonumber
z & = & \biggl( \tilde Q, \tilde s \biggr) \\
\nonumber
  & = & \biggl( \tilde Q, \tilde n \biggr) + 
    \biggl( \tilde Q, {D \over d} \bigl( \cos \phi \tilde T_c + 
           \sin \phi \tilde T_s\bigr) {\rm e}^{-2\pi i f t_0} \biggr).
\end{eqnarray}
Its expectation value is the complex number
\begin{equation}
\langle z \rangle = {D \over d} \bigl(\cos \phi + i \sin \phi \bigr) = {D \over d} {\rm e}^{i \phi}.
\end{equation}
The modulus of this complex number is the (expected) inverse distance,
and its phase is the (expected) phase.  Note that because the
normalization of $\tilde Q$ has changed, the expected value
\begin{equation}
\langle |z|^2 \rangle = 2 + \bigl({D \over d} \bigr)^2.
\end{equation}
is larger than in the single phase case.  The additional uncertainty
about the phase $\phi$ means that the distance to the source can not
be determined as accurately as in the single phase case.  Following
conventional practice in the field, the modulus $|z|$ will be called
the ``Signal to Noise Ratio'' (SNR) although since in the absence of a
source its mean-square value is two, one might argue that
$|z|/\sqrt{2}$ is the quantity that should carry this name.

To construct the $\chi^2$ statistic, we choose frequency bands as
before.  We will assume that $\tilde T_c$ and $\tilde T_s$ have {\it
identical} frequency bands and are orthogonal in each of these
bands \footnote{ Even if the templates are not orthogonal in each of
the $p$ bands, it is possible to apply the Gram-Schmidt procedure in
{\it each} of these bands separately.  But the subsequent template
would no longer be an optimal filter for the desired signal.}.  This
is exactly true in the stationary-phase approximation where $\tilde
T_s(f) = i \tilde T_c(f)$ for $f > 0$ and $\tilde T_s(f) = - i \tilde
T_c(f)$ for $f < 0$. Thus
\begin{eqnarray} \bigl( \tilde T_c, \tilde T_c \bigr)_j & = & \bigl(
\tilde T_s, \tilde T_s \bigr)_j = { 1 \over p}, \text{ and} \\ \bigl(
\tilde T_c, \tilde T_s \bigr)_j & = & 0.
\end{eqnarray}
We define the complex signal $z_j$ in the $j$'th band as before
\[
z_j  \equiv  \bigl( \tilde Q, \tilde s \bigr)_j
\]
and also define $\Delta z_j$ as before
\begin{equation}
\Delta z_j \equiv z_j - { z \over p}.
\end{equation}
One then finds
\begin{eqnarray}
\nonumber
\langle z_j \rangle & = & {1 \over p} {D \over d} {\rm e}^{i \phi} \cos \theta \\
\nonumber
\langle | z_j |^2 \rangle & = & {2 \over p} + {1 \over p^2} \biggl( {D \over d} \biggr)^2\\
\nonumber
\langle z_j^* z  \rangle  & = & {2 \over p} + {1 \over p }  \biggl( {D \over d} \biggr)^2\\
\label{e:thisone}
\langle | \Delta z_j |^2 \rangle & = & {2 \over p}\bigl( 1 - {1 \over p} \bigr).
\end{eqnarray}
The $\chi^2$ statistic is defined by \footnote{In early versions of
GRASP (prior to April 1998) the author mistakenly thought that this
two-phase quantity could be calculated by substituting the template
defined by the maximum-likelihood estimator of $\phi$ into the
known-phase formula (\ref{e:definechisquare}). This was corrected in
GRASP by Allen, Brady, and Creighton in June 1998. Note that if one
{\it fixes} the value of $\phi$ then for Gaussian noise one obtains a
statistic with a $\chi^2$ distribution and $p-1$ degrees of
freedom. But if $\phi$ is estimated from the data then it depends (in
a non-linear way) on the detector noise, and so even for Gaussian
noise, the resulting statistic does {\it not} have a $\chi^2$
distribution.}
\begin{equation}
\label{e:twophasedef}
\chi^2 = p \sum_{j=1}^p | \Delta z_j |^2
\end{equation}
and thus from (\ref{e:thisone}) has expected value
\begin{equation}
\langle \chi^2 \rangle = 2p - 2.
\end{equation}
In Appendix~\ref{s:appendix} we show that if the detector noise is
Gaussian, then $\chi^2$ has a classical $\chi^2$ probability
distribution.  Because both the real {\it and} imaginary parts of
$\Delta z_j$ sum to zero, the number of (real) degrees of freedom is
$2p -2$.

We now consider the case where the astrophysical waveform $h(t) = {D'
\over d'} T'(t)$ does {\it not} exactly match any linear combination
of the templates $T_c$ and $T_s$.  This is to be expected from real
signals if the templates form a discrete finite grid in parameter
space.  As before, with no loss of generality we assume that $t_0=0$
and that $D'$ is chosen so that
\[
\bigl( \tilde T' , \tilde T' \bigr) = 1.
\]
Consider the possible values, as $\psi \in [0, 2\pi)$ varies, of the inner product
\[
\bigl( \cos \psi \tilde T_c + \sin \psi \tilde T_s , \tilde T').
\]
Since both $\cos \psi \tilde T_c + \sin \psi \tilde T_s$ and $\tilde
T'$ are unit length, Schwartz's inequality implies that this inner
product lies in the range $[-1,1]$, and its maximum value must lie in
the range $[0,1]$. This maximum value (see Appendix~\ref{s:max}) is
\begin{equation}
\cos \theta \equiv \sqrt{ \bigl( \tilde T_c , \tilde T')^2 + \bigl( \tilde T_s , \tilde T')^2 },
\end{equation}
which defines $\theta \in [0, \pi/2]$.  This in turn defines the
mismatch $\epsilon = 1 - \cos \theta$ between the template $T'$ and
the one-parameter family of templates. Note that (in contrast to the
single-phase case) the maximization over $\psi$ {\it automatically}
leads to $\cos \theta \in [0,1]$ and hence $\epsilon \in [0,1]$.  We
can also define $\phi \in [0, 2 \pi)$ by
\begin{eqnarray}
\nonumber
\cos \phi \cos \theta & \equiv & \bigl( \tilde T_c , \tilde T' \bigr), \text{ and}\\
\label{e:lastminute}
\sin \phi \cos \theta & \equiv & \bigl( \tilde T_s , \tilde T' \bigr).
\end{eqnarray}
Thus one has
\begin{equation}
\label{e:definephitheta}
\bigl( \tilde T_c + i \tilde T_s , \tilde T' \bigr) = {\rm e}^{i \phi} \cos \theta.
\end{equation}
This equation may be taken as the {\it definition} of $\phi$ and
$\theta$.

The filter output is given by
\[
z = \bigl(\tilde Q, \tilde s) = \bigl(\tilde Q, \tilde n + \tilde h)
\]
and thus has expectation value
\begin{eqnarray}
\nonumber
\langle z \rangle & = & {D' \over d'} \bigl(\tilde T_c + i \tilde T_s, \tilde T' \bigr) \\
& = &  {D' \over d'} {\rm e}^{i \phi} \cos \theta.
\end{eqnarray}
The expected square modulus of the filter output is
\[
\langle | z |^2 \rangle = 2 + \biggl(  {D' \over d'} \biggr)^2 \cos^2 \theta = 2 + | \langle z \rangle |^2.
\]
We now investigate the effect of the template/signal mismatch on the
$\chi^2$ statistic.

To begin, we need to characterize the overlap between the signal and
the templates in the $j$'th frequency band.  Define complex quantities
$\lambda_j$ by
\begin{equation}
\bigl(\tilde T_c + i \tilde T_s , \tilde T' \bigr)_j = \lambda_j {\rm e}^{i \phi} \cos \theta.
\end{equation}
Using (\ref{e:definephitheta}), these complex quantities are
constrained by
\begin{equation}
\sum_{j=1}^p \lambda_j = 1.
\end{equation}
The filter output in the $j$'th frequency band is given by
\[
z_j = \bigl( \tilde T_c + i \tilde T_s , \tilde n \bigr)_j + {D' \over d'} \lambda_j {\rm e}^{i \phi}  \cos \theta.
\]
and the various expectation values in the $j$'th band are
\begin{eqnarray}
\nonumber
\langle z_j \rangle & = &  {D' \over d'} \lambda_j {\rm e}^{i \phi}\cos \theta \\
\nonumber
\langle | z_j |^2 \rangle & = & {2 \over p} +  \biggl( {D' \over d'}\biggr)^2 |\lambda_j |^2 \cos^2 \theta \\
\nonumber
\langle z_j^* z  \rangle  & = & {2 \over p} + \biggl( {D' \over d'}\biggr)^2 \lambda^*_j \cos^2 \theta \\
\nonumber
\langle | \Delta z_j |^2 \rangle & = & {2 \over p}\bigl( 1 - {1 \over p} \bigr) + 
 \biggl( {D' \over d'}\biggr)^2 \bigl| \lambda_j - {1 \over p} \bigr|^2  \cos^2 \theta
\end{eqnarray}
The expected value of $\chi^2$ is then
\begin{equation}
\label{e:mainresult3}
\langle \chi^2 \rangle = 2p - 2 + \kappa | \langle z \rangle |^2 
\end{equation}
with
\begin{equation}
\label{e:kappa2}
\kappa \equiv p \sum_{j=1}^p \biggl| \lambda_j - {1 \over p}  \biggr|^2 = -1 + p\sum_{j=1}^p \bigl| \lambda^2_j \bigr|.
\end{equation}
To place an upper limit on $\kappa$, note that from Schwartz's
inequality, for {\it any} value of the angle $\psi$, one has
\footnote{In these inequalities, the quantities that appear on both
sides are real because $T_c$, $T_s$, and $T'$ are all real in the time
domain.  Thus in the frequency domain they obey $\tilde Z(f)=\tilde
Z^*(-f)$.  Therefore if we let $X$ and $Y$ denote {\it any} of $T_c$,
$T_s$, and $T'$, the nine possible inner products $\bigl( \tilde X,
\tilde Y \bigr)$ are all real.}
\begin{eqnarray}
\nonumber
\bigl( \cos \psi \tilde T_c + \sin \psi \tilde T_s \; , \; \tilde T' \bigr)_j^2
& \le &
{1 \over p} \bigl(\tilde T' , \tilde T' \bigr)_j \\
\nonumber
\bigl( \cos \psi \bigl( \tilde T_c, \tilde T' \bigr)_j + \sin \psi \bigl( \tilde T_s, \tilde T' \bigr)_j \bigr)^2 & \le & 
{1 \over p} \bigl(\tilde T' , \tilde T' \bigr)_j.
\end{eqnarray}
The {\it maximum} value of the left-hand-side (see
Appendix~\ref{s:max}) is
\begin{eqnarray}
\nonumber
&& \bigl( \tilde T_c, \tilde T' \bigr)^2_j + \bigl( \tilde T_s, \tilde T' \bigr)^2_j = \\
\nonumber
&& \bigl( \Re \bigl[ {\lambda_j {\rm e}^{i \phi}} \bigr] \cos \theta \bigr)^2 + 
\bigl( \Im \bigl[ {\lambda_j {\rm e}^{i \phi}} \bigr] \cos \theta \bigr)^2 =   \\
\nonumber
&& |\lambda_j|^2 \cos^2 \theta,
\end{eqnarray}
where we have made use of (\ref{e:lastminute}) and
(\ref{e:definephitheta}).  Thus
\[
 \bigl| \lambda_j \bigr|^2  \le {1 \over p\cos^2 \theta } \bigl(\tilde T' , \tilde T' \bigr)_j.
\]
Summing both sides over $j$ and making use of (\ref{e:kappa2}) this
implies that
\begin{equation}
\label{e:kappalimit4}
0 \; \le \; \kappa \; \le \; {1 \over \cos^2 \theta} -1.
\end{equation}
This result makes no assumptions about the form of the mismatched
signal $\tilde T'$.  As in the single phase case, if we assume that
the mismatch is small, $\epsilon \ll 1$, we obtain
\begin{equation}
\label{e:kappalimit4e}
0 \; \le \; \kappa \; \le \; 2 \epsilon .
\end{equation}
This result does not assume any relationship between the frequency
bands of the signals $T$ and $T'$.

If we assume that the bands $\Delta f_j$ for the mismatched signal
$T'$ are the {\it same} as those for the templates $T_c$ and $T_s$
then we can obtain a much stronger upper bound. For example, this is
the case if all three templates are drawn from a family of
stationary-phase approximate inspiral chirps. In this case, $(\tilde
T', \tilde T')_j = 1/p$ and the same logic as in the single-phase case
can be used to establish that
\begin{equation}
\label{e:kappalimit3}
\kappa \le (p-1) \biggl( { 1 \over \cos \theta} -1 \biggr)^2 \sim (p-1) \epsilon^2,
\end{equation}
provided that $p>2$ and $\epsilon = 1 - \cos \theta \le 2/p$.
Thus, with some minor modifications, all the single-phase results
apply to the unknown phase case.

In the case where the signal and template are not a perfect match, the
expected value of $\Delta z_j$ does not vanish:
\[
\langle \Delta z_j \rangle = {D' \over d'} {\rm e}^{i \phi} \; \cos
\theta \;\bigl( \lambda_j - {1 \over p} \bigr).
\]
If the detector noise is stationary and Gaussian, then for a given
astrophysical signal $T'$ and filter-template $T_{c,s}$ this means
that the probability distribution of $\chi^2$ is a classical
non-central $\chi^2$ distribution \footnote{The fact that template
mismatch leads to a {\it non-central} $\chi^2$ distribution for
Gaussian detector noise was first pointed out by Jolien Creighton.}
with $2p-2$ degrees of freedom and non-centrality parameter
\[
\lambda= p \sum_{j=1}^p | \langle \Delta z_j \rangle |^2 = \kappa | \langle z
\rangle|^2.
\]
Unfortunately, for a set of candidate events, which correspond to
different waveforms $D' T'/d'$ and ring off different templates $T$,
the values of $\kappa$ have different, unknown values, bounded only by
(\ref{e:kappalimit4}) or (\ref{e:kappalimit3}).  In this case, since
the average of non-central $\chi^2$ distributions with different
values of $\lambda$ is {\it not} a non-central $\chi^2$ distribution,
one can only bound the expected distribution, not determine it from
first principles.

\section{Thresholding Conditions}
\label{s:thresholds}

As described in Section~\ref{s:intro}, the $\chi^2$ time-frequency
discriminator is most often used as a veto.  For a given data set, the
threshold value $\chi^2_*$ is usually determined using Monte-Carlo
simulation of signals, analytic guidance, and experience.  If the
signal and template were known to have identical form, then the
threshold $\chi^2_*$ would be a number.  However since the signal and
template are not expected to match perfectly, the threshold $\chi^2_*$
is a function of the observed SNR.

It is helpful to understand the threshold that would be appropriate
for stationary Gaussian noise.  In this case, the optimal threshold is
given by the inverse of the non-central $\chi^2$ cumulative
distribution function \footnote{For example the {\bf Matlab} function
ncx2inv(r,2p-2,$\lambda$), where $r$ is the probability that a signal
in Gaussian noise would lie below this threshold, for example,
99.9\%.}.  A putative signal whose $\chi^2$ value is smaller than this
``Gaussian noise'' threshold is likely to merit further examination,
even if the noise is {\it not} Gaussian
\footnote{The converse is not true: with non-Gaussian noise, true
signals may well have $\chi^2$ values that exceed a reasonable
threshold for the Gaussian case.}.  So in most cases a reasonable
threshold will be greater than or equal to the threshold appropriate
for Gaussian detector noise.

\begin{figure}
\begin{center}
\epsfig{file=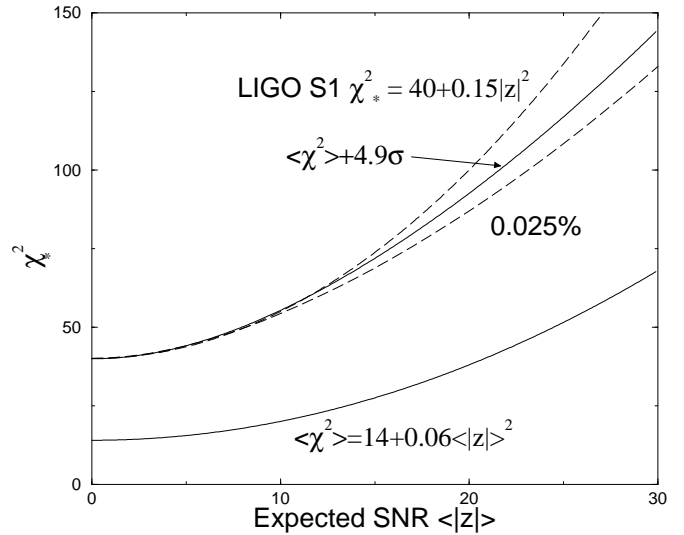,angle=-90,width=3.4in}
\caption{ \label{f:comparison} A comparison of different thresholds
$\chi^2_*$ for the $\chi^2$ time-frequency discriminator, for the LIGO
S1 two-phase case with $p=8$, $\epsilon=0.03$, and worst-case
$\kappa=2\epsilon$.  The bottom solid line shows the expected value of
$\chi^2$ for stationary Gaussian noise given by (\ref{e:meanchi2}).
The top solid line shows a threshold (\ref{e:athresh}) set 4.9
$\sigma$ above this expected value.  The upper dashed line shows the
heuristic threshold (\ref{e:ligoS1threshold}) used for the LIGO S1
analysis, which was determined from Monte-Carlo studies. The lower
dashed line shows the threshold which would be exceeded with
probability 0.025\% in stationary Gaussian noise.}
\end{center}
\end{figure}

In the stationary Gaussian case, for fixed $T$ and $T'$, the expected
value, variance $\sigma^2$, and standard deviation $\sigma$ of the
non-central $\chi^2$ distribution with $2p-2$ degrees of freedom are
\cite{stats}
\begin{eqnarray}
\label{e:meanchi2}
\langle \chi^2 \rangle & = & 2p - 2 + \lambda = 2p-2 + \kappa |\langle z \rangle|^2,\\
\nonumber
\sigma^2 & = & \langle (\chi^2)^2 \rangle - \langle \chi^2 \rangle^2 =  4 p - 4 + 4\lambda, 
                 \text{ and}\\
\nonumber
\sigma & = & \sqrt{4 p - 4 + 4 \kappa |\langle z \rangle|^2}.
\end{eqnarray}
In the neighborhood of the maximum, for values of the non-centrality
parameter $\lambda$ significantly larger than $2p-2$, the non-central
$\chi^2$ distribution is approximately a Gaussian of width $\sigma$,
centered about the mean value $2p - 2 + \lambda$.  In this case, the
optimal $\chi^2$ veto threshold for Gaussian noise is
well-approximated by
\[
\chi^2_* = \langle \chi^2 \rangle + \text{few } \sigma,
\]
where $\sigma$ are the expected statistical fluctuations in $\chi^2$
evaluated for the ``worst-case'' value of $\kappa$.  If we assume that
the putative signals and templates do not share the same frequency
bands $\Delta f_j$, so that the upper limit of (\ref{e:kappalimit4})
applies, then we obtain a $\chi^2$ threshold of the form
\begin{equation}
\label{e:gaussianthreshold}
\chi^2_* = 2 p -2 + 2 \epsilon |\text{SNR}|^2 + \text{few } \sqrt{4 p - 4 + 8 \epsilon
|\text{SNR}|^2},
\end{equation}
where we have replaced the expected SNR by the measured SNR
\footnote{It is reasonable to make this replacement, since the case of
interest is large SNR, and for this case the fractional statistical
fluctuations in the SNR are small, so that $\text{SNR} \approx \langle
\text{SNR} \rangle$.}.  Although we have justified this approximation
to the threshold for large non-centrality parameter $\lambda$, it
turns out to be a reasonably good approximation even when the
non-centrality parameter $\lambda$ is small
\footnote{An even better approximation can be found in \cite{as}.  This leads to a $\chi^2$ threshold of
$$ \langle \chi^2 \rangle + b(x_p^2-1) + 2 x_p \sqrt{b(\langle \chi^2
\rangle - b)}$$ where $b={\kappa \langle z \rangle^2 + p -1 \over
\kappa \langle z \rangle^2 + 2 p -2}$ increases smoothly over the range
$[1/2,1]$ as the SNR increases, and $x_p$ is the corresponding
probability threshold for a zero mean unit variance Gaussian.}.

It is interesting to compare the threshold appropriate for Gaussian
noise to the $\chi^2$ threshold used in the LIGO S1 analysis for $p=8$
and $\epsilon=0.03$, which is equation (4.7) of reference
\cite{ligo_inspiral}
\begin{equation}
\label{e:ligoS1threshold}
\text{LIGO S1 }\chi^2_* = 40 + 0.15 |\text{SNR}|^2,
\end{equation}
shown as the upper dashed line in Figure~\ref{f:comparison}. The lower
solid curve shows the expectation value of $\chi^2$ given by
(\ref{e:mainresult3}), and the upper solid curve shows a 4.9 standard
deviation threshold given by (\ref{e:gaussianthreshold}), which is
\begin{equation}
\label{e:athresh}
\chi^2_*= 14 + 0.06 |\text{SNR}|^2 + 4.9 \sqrt{28 + 0.24 |\text{SNR}|^2}.
\end{equation}
As is clear from the graph, the heuristic threshold is reasonably well
matched by the sort of threshold that one might set based on a
worst-case analysis for Gaussian detector noise. 

For {\it very} large SNR, the Gaussian threshold condition
(\ref{e:gaussianthreshold}) consists of two terms.  The dominant term
(quadratic in SNR) comes from the mean value $\langle \chi^2 \rangle$
and has coefficient exactly $\kappa$.  The sub-dominant term (linear
in SNR) comes from a few times $\sigma$. Hence, a threshold like the
LIGO S1 choice would {\it not} veto high SNR events that {\it could}
be confidently vetoed in Gaussian noise.  This is illustrated in
Figure~\ref{f:largesnr}.

In the LIGO S1 analysis, which sets an upper limit on the Galactic
inspiral rate, the probability of observing a close inspiral (very
large SNR) is far smaller than the probability of observing a more
distant (low SNR) event from near the Galactic center.  Thus, applying
the more stringent thresholding condition at high SNR would probably
not have had a significant detrimental effect on the analysis: it
would not have significantly decreased the detection efficiency.
However it also would not have improved the analysis, since the
highest SNR events that passed the $\chi^2$ threshold had SNR less
than 16.

\begin{figure}
\begin{center}
\epsfig{file=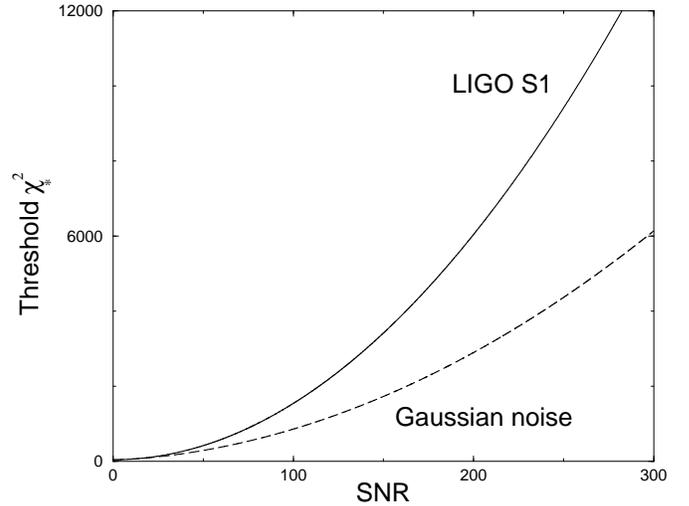,angle=-90,width=3.4in}
\caption{ \label{f:largesnr} A comparison of different thresholds for
 the $\chi^2$ time-frequency discriminator, at large SNR.  The upper
 solid curve is the LIGO S1 threshold (\ref{e:ligoS1threshold}).  The
 lower (dashed) curve is the threshold that would be exceeded with
 probability 0.025\% in stationary Gaussian noise.  (On the scale of
 this graph the approximation (\ref{e:athresh}) to the dashed curve is
 difficult to distinguish from the exact result obtained from the
 inverse of the non-central $\chi^2$ distribution.)}
\end{center}
\end{figure}

\section{Unequal expected SNR Frequency Intervals}
\label{s:unequalintervals}
One may also define a $\chi^2$ time-frequency discriminator using
frequency intervals which do {\it not} make equal expected
contributions to the SNR, and almost all of the previous results hold
\footnote{Stanislav Babak pointed out that this case was of interest,
because for practical reasons, it may not be possible to construct
``equal SNR'' intervals.}. For simplicity, in this Section we treat
only the single-phase case.

Begin by making the same initial assumptions as in
Section~\ref{s:mismatch}, but choose frequency intervals which do {\it
not} make equal expected contributions to the SNR. Thus
\begin{equation}
\label{e:unequalintervals}
\text{choose } \Delta f_j \text{ so that }
\bigl( \tilde T , \tilde T)_j = q_j,
\end{equation}
where the $q_j \in (0,1]$ do not necessarily equal $1/p$. Template
normalization $( \tilde T , \tilde T) = 1$ implies that they satisfy
$\sum_j q_j = 1$.

Define the SNR in the $j$'th band as previously
\begin{equation}
z_j  = (\tilde Q, \tilde s)_j,
\end{equation}
and define $\Delta z_j$ as the difference between observed SNR in the
$j$'th band, and the value that would be {\it anticipated}
\footnote{As previously, the word ``anticipated'' is used, rather than
``expected'', because we do not experimentally or observationally have
access to the expected value of the SNR $\langle z \rangle$.  In other
words, the quantities $\Delta z_j = z_j - q_j z $ are {\it not} the
same as $\delta z_j = z_j - \langle z_j \rangle = z_j - q_j \langle z
\rangle$.  For any given observation or data set, the quantities
$\delta z_j$ do {\it not} sum to zero (although their expectation
value does).  They have expected square values $\langle (\delta z_j
)^2 \rangle = q_j$.}  based on the total SNR observed.  Thus we define
\begin{equation}
\Delta z_j \equiv z_j - q_j z,
\end{equation}
where as before the observed SNR is $z = \sum_j z_j = (\tilde Q,
\tilde s)$.  Note that by definition the sum $\sum_j \Delta z_j$
vanishes.

Consider the case of a signal waveform $T'$ which may be mis-matched
to the template $T$, so $(T,T') = \cos \theta$ and (without loss of
generality) $0 \le \cos \theta \le 1$.  Within the $j$'th band, the template
$T'$ has overlap
\begin{equation}
\bigl( \tilde T, \tilde T' \bigr)_j = \lambda_j \cos \theta,
\end{equation}
which may be taken to define the quantities $\lambda_j$.
As before, the sum $\sum_j \lambda_j = 1$.  Taking the signal to be
\begin{equation}
 s(t) = n(t) + {D' \over d'} \; T'(t),
\end{equation}
we can now compute the various expectation values.

It follows immediately from the above definitions that these
expectation values are given by
\begin{eqnarray}
\nonumber
\langle z \rangle & = & {D' \over d'} \cos \theta, \\
\nonumber
\langle z_j \rangle & = & {D' \over d'} \lambda_j \cos \theta, \\
\nonumber
\langle \Delta z_j \rangle & = & {D' \over d'} (\lambda_j - q_j) \cos \theta, \\
\nonumber
\langle z^2 \rangle & = & 1 +  \biggl( {D' \over d'} \biggr)^2  \cos^2 \theta, \\
\nonumber
\langle   z_j z_k \rangle & = & q_j \delta_{jk} +  \biggl( {D' \over d'} \biggr)^2 \lambda_j \lambda_k \cos^2 \theta, \\
\nonumber
\langle z_j^2 \rangle & = & q_j +  \biggl( {D' \over d'} \biggr)^2 \lambda^2_j \cos^2 \theta, \text{ and} \\
\nonumber
\langle (\Delta z_j)^2 \rangle & = & q_j (1-q_j) +  \biggl( {D' \over d'} \biggr)^2 (\lambda_j - q_j)^2 \cos^2 \theta, \\
\label{e:deltazform}
\end{eqnarray}
where (the Kronecker symbol) $\delta_{jk}=1$ if $j=k$ and vanishes
otherwise.  The formulae of Section~\ref{s:mismatch} correspond to the
special case in which $q_j = 1/p$.

Define the $\chi^2$ time-frequency discriminator in the unequal
expected SNR interval case to be
\begin{equation}
\label{e:anotherdef}
\chi^2 = \sum_{j=1}^p (\Delta z_j)^2/q_j.
\end{equation}
To characterize the statistical properties of $\chi^2$, it is helpful
to express it in terms of a different set of variables.  Begin by
writing $\chi^2$ as
\begin{eqnarray}
\nonumber
\chi^2  & = & \sum_{j=1}^p (z_j - q_j z) \Delta z_j / q_j \\
\nonumber 
        & = & \sum_{j=1}^p z_j \Delta z_j /q_j \\
        & = & \sum_{j=1}^p  z^2_j /q_j  - \biggl(\sum_{j=1}^p  z_j \biggr)^2,
\end{eqnarray}
where we have made use of the fact that the $\Delta z_j$ sum to zero.
Define new variables $u_j = z_j/\sqrt{q_j}$, which have variance unity
and are uncorrelated:
\[
\langle u_j u_k \rangle - \langle u_j \rangle  \langle u_k\rangle = \delta_{jk},
\]
In terms of these variables, the statistic is
\begin{equation}
\label{e:firstform}
\chi^2 = \sum_{j=1}^p u_j^2 - \biggl(\sum_{j=1}^p  \sqrt{q_j} u_j \biggr)^2.
\end{equation}
To characterize the probability distribution of $\chi^2$, it is
convenient to change variables once again.

We introduce new variables $v_j$ which are linear combinations of the
$u_j$ and are most conveniently written in matrix form as
\begin{equation}
\label{e:orthotransform}
\left[ \matrix{
                   v_1 & \cr
                   \vdots & \cr
                   v_p & \cr }
\right] = M \left[ \matrix{
                   u_1 & \cr
                   \vdots & \cr
                   u_p & \cr }
\right]
\end{equation}
where $M$ is a $p \times p$ square matrix.  Choose $M$ to be an
orthogonal matrix, which thus satisfies $M^t M = M M^t = I$, where $t$
denotes transpose and $I$ is the $p\times p$ identity matrix.  These
linear transformations (\ref{e:orthotransform}) are rotations
\footnote{ The $p(p-1)/2$-dimensional orthogonal group O(p) consists
of two disconnected components.  In one of these components, the
determinants of the matrices $M$ are $+1$ and in the other component
the determinants are $-1$. The first component contains the identity
matrix $I$. Strictly speaking, only the matrices in the component
connected to the identity may be thought of as pure rotations.
Matrices in the other component are combinations of pure rotations and
reflections that do not preserve the ``handedness'' or orientation of
the basis.} that map $\mathbb{R}^p \to \mathbb{R}^p$.

Since the transformation is orthogonal, the new variables $v_j$ also
have variance unity and are uncorrelated:
\begin{eqnarray}
\nonumber
\langle v_j v_k \rangle & - & \langle v_j \rangle  \langle v_k\rangle  \\ 
\nonumber
& = & \sum_{\ell=1}^p \sum_{m=1}^p M_{j \ell} M_{k m} \bigl[
\langle u_\ell u_m \rangle - \langle u_\ell \rangle  \langle u_m\rangle \bigr] \\
\nonumber
& = & \sum_\ell \sum_m M_{j \ell} M_{k m} \delta_{\ell m} \\
\nonumber
& = & \sum_m M_{j m} M_{k m} \\ 
\nonumber
& = & \sum_m M_{j m} (M^t)_{m k} \\ 
& = & (M M^t)_{jk} = I_{jk}=\delta_{jk}.
\end{eqnarray}
Moreover
\begin{equation}
\label{e:preserveslength}
\sum_{j=1}^p u_j^2 = \sum_{j=1}^p v_j^2,
\end{equation}
since rotations do not change the length of a vector.

The rotation $M$ may be chosen so that any given orthonormal basis is
mapped onto any other orthonormal basis of the same orientation
(handedness).  Thus one may choose the rotation so that the last of
the new variables is
\begin{equation}
\label{e:lastv}
v_p = \sqrt{q_1} u_1 + \sqrt{q_2} u_2 + \cdots \sqrt{q_p} u_p.
\end{equation}
This corresponds to constraining the final row of $M$ to be
$(\sqrt{q_1}, \cdots, \sqrt{q_n})$, or equivalently to requiring that
$M$ map the vector on the l.h.s. below to the final ($p$'th) basis
vector.
\[
\left[ \matrix{
                   \sqrt{q_1} & \cr
                   \vdots & \cr
                   \sqrt{q_n} & \cr }
\right] 
\matrix{
                   M  & \cr
                   \longrightarrow & \cr }
\left[ \matrix{
                   0 & \cr
                   \vdots & \cr
                   0 & \cr
                   1 & \cr }
\right]
\]
In terms of these new variables, $\chi^2$ given in (\ref{e:firstform})
may be written using (\ref{e:preserveslength}) and (\ref{e:lastv}) as
\begin{equation}
\label{e:pminus1dof}
\chi^2 = \biggl( \sum_{j=1}^{p} v_j^2 \biggr) - v_p^2 =
\sum_{j=1}^{p-1} v_j^2.
\end{equation}
This form makes it easy to characterize the statistics of $\chi^2$ if
the detector noise is Gaussian \footnote{Another approach to showing
that (\ref{e:anotherdef}) has a non-central $\chi^2$ distribution for
Gaussian detector noise is to write it as
$$
\chi^2 = \sum_{j=1}^p \sum_{k=1}^p {z_j \over \sqrt{q_j}} \; P_{jk} { z_k \over \sqrt{q_k}},
$$ where $P$ is a $p \times p$ matrix.  In the single-phase
(two-phase) case, $P$ may be taken to be real and symmetric
(Hermitian).  One need only show that $P$ has rank $p-1$ and is a
projection operator, meaning that $P^2 = P$.  This is equivalent to
saying that $P$ has one zero eigenvalue and that all its {\it other}
eigenvalues are equal to one.  This implies that an orthogonal
(unitary) transformation can be found which puts $P$ into block
diagonal form with a one sub-block proportional to the $p-1$
dimensional identity matrix and the other sub-block vanishing.  It
follows that for Gaussian detector noise $\chi^2$ has a classical
non-central $\chi^2$ distribution with $p-1$ ($2p-2$) degrees of
freedom.}.

If the noise is Gaussian, then the $u_j$ are uncorrelated (and hence
independent) Gaussian random variables with unit variance, and the
$v_k$ are also uncorrelated (and hence independent) Gaussian random
variables with unit variance.  Thus, in the case of Gaussian noise it
immediately follows from (\ref{e:pminus1dof}) that $\chi^2$ has a
classical non-central $\chi^2$ distribution with $p-1$ degrees of
freedom.  In the two-phase case, each of the $v_j$ is a complex
variable with independent real and imaginary parts and the resulting
distribution has $2p-2$ degrees of freedom.

The non-centrality parameter $\lambda$ may be evaluated by calculating
the expected value of $\chi^2$.  Using (\ref{e:deltazform}) and
(\ref{e:anotherdef}) this is
\begin{equation}
\langle \chi^2 \rangle = p - 1 + \kappa \biggl( {D' \over d'} \biggr)^2
\cos^2 \theta = p - 1 + \kappa \langle z \rangle^2,
\end{equation}
and hence the non-centrality parameter is given by $\lambda = \kappa
\langle z \rangle^2 $. The constant $\kappa$, which is determined by
the choice of intervals, the spectrum of the detector noise, and the
frequency-dependence of the mismatch between templates, is given by
\begin{eqnarray}
\nonumber
\kappa & = & \sum_{j=1}^p (\lambda_j - q_j)^2/q_j \\
\nonumber
       & = & \sum_{j=1}^p \bigl( \lambda_j^2/q_j - 2 \lambda_j+q_j \bigr) \\
\label{e:lambdamod}
       & = & -1 + \sum_{j=1}^p \lambda_j^2/q_j.
\end{eqnarray}
Clearly $\kappa$ is non-negative.  One can easily obtain an upper
limit on $\kappa$ even in this case where the frequency intervals are
not ``equal SNR'' intervals.

To obtain a limit on $\kappa$, begin with Schwartz's inequality, which
implies that
\begin{eqnarray}
\nonumber
 \bigl( \tilde T, \tilde T' \bigr)_j^2 & \le &   \bigl( \tilde T, \tilde T \bigr)_j  \bigl( \tilde T', \tilde T' \bigr)_j \\
\nonumber
\lambda^2_j \cos^2 \theta & \le &  q_j \bigl( \tilde T', \tilde T' \bigr)_j \\
\nonumber
& \Rightarrow &  \\
\lambda_j^2/q_j  & \le & {1 \over \cos^2 \theta } \bigl( \tilde T', \tilde T' \bigr)_j.
\end{eqnarray}
Summing both sides over $j$ and using (\ref{e:lambdamod}) yields
\begin{equation}
0 \; \le \; \kappa \; \le \; {1 \over \cos^2 \theta} -1,
\end{equation}
and hence for $\epsilon \ll 1$ one has $0 \le \kappa \le 2 \epsilon$,
just as in the ``equal SNR interval'' case.

One can also establish a stronger limit analogous to
(\ref{e:kappalimit2}) for the case where the templates $T$ and $T'$
have the same values of $q_j$ for a given set of frequency intervals.
In this case, Schwartz's inequality implies that
\begin{eqnarray}
\nonumber
 \bigl( \tilde T, \tilde T' \bigr)_j^2 & \le &  \bigl( \tilde T, \tilde T \bigr)_j  \bigl( \tilde T', \tilde T' \bigr)_j \\
\nonumber
\lambda^2_j \cos^2 \theta & \le &  q^2_j,
\end{eqnarray}
and hence that
\begin{equation}
\label{e:funkyconst}
-{ q_j \over \cos \theta} \le \lambda_j \le  { q_j \over \cos \theta}.
\end{equation}
Without loss of generality, relabel the frequency intervals so that
\[q_1 \le q_2 \le \cdots \le q_p.
\]
The value of $\kappa$ is maximized by setting:
\begin{eqnarray}
\nonumber
\lambda_1 & = & 1+ (q_1 - 1)/\cos \theta \\
\nonumber
\lambda_2 & = & q_2/\cos \theta \\
\nonumber
& \cdots & \\
\label{e:funkyvals}
\lambda_p & = & q_p/\cos \theta .
\end{eqnarray}
This choice satisfies the constraint that $\sum_j \lambda_j = 1$, and
the r.h.s. of (\ref{e:funkyconst}).  In order that $\lambda_1$ satisfy
the l.h.s. of the constraint (\ref{e:funkyconst}) we need to have
\begin{equation}
1 - \cos\theta < 2 q_1
\end{equation}
or equivalently $\epsilon < 2 q_1$. For the values of $\lambda_j$
given in (\ref{e:funkyvals}) one then obtains
\begin{eqnarray}
\nonumber
\kappa & = &  \sum_{j=1}^p (\lambda_j -  q_j)^2/q_j \\
\nonumber
       & = & \bigl((q_1 - 1)/\cos \theta + 1 - q_1 \bigr)^2/q_1 + \\ 
\nonumber
       &    &q_2 \bigl( {1 \over \cos \theta} -1 \bigr)^2 + \cdots + q_p \bigl( {1 \over \cos \theta} -1 \bigr)^2 \\
\nonumber
       & =  & \bigl[ (q_1 - 1)^2/q_1  + q_2 + \cdots + q_p \bigr] \bigl( {1 \over \cos \theta} -1 \bigr)^2  \\
\nonumber
       & = &  \bigl[ {1 \over q_1} -1  \bigr] \bigl( {1 \over \cos \theta} -1 \bigr)^2 .
\end{eqnarray}
For $p>2$ this gives the upper bound on $\kappa$.

Denote the smallest value of $q_j$ ($q_1$ just above) by $q_{\rm
min}$.  Then, in the case where the templates $T$ and $T'$ have the
same values of $q_j$ for a given set of frequency intervals, and $1 -
\cos\theta < 2 q_{\rm min}$ one has
\begin{equation}
0 \le \kappa \le \biggl( {1 \over q_{\rm min}} -1  \biggr) \biggl( {1 - \cos \theta \over \cos \theta} \biggr)^2.
\end{equation}
If $\epsilon = 1 - \cos \theta$ is much less than unity, then this may be written
\begin{equation}
0 \le \kappa \le \biggl( {1 \over q_{\rm min}} -1  \biggr) \epsilon^2 \text{ if } \epsilon < 2 q_{\rm min}.
\end{equation}
These reduce to the previous results of Section~\ref{s:mismatch} when
all the $q_j$ (and hence $q_{\rm min}$) equal $1/p$.

This ``unequal expected SNR'' $\chi^2$ discriminator may be of
practical use when it is impossible to construct equal SNR intervals.
It may also permit the construction of discriminators which are
specifically tuned to common types of detector noise.

\section{Other types of $\chi^2$ tests}
\label{s:othertests}

There are many possible $\chi^2$ tests that could be used to
discriminate spurious signals from genuine ones.  Here we compare the
$\chi^2$ time-frequency discriminator of this work with a standard 
$\chi^2$ test used by Baggio et. al. \cite{baggioetal}, which we
denote by $\bar \chi^2$.  This tests ``goodness of fit'' for a modeled
signal embedded in stationary Gaussian noise, and is used in analyzing
data from AURIGA, a narrow-band resonant-bar gravitational wave
detector.

It is instructive to express this standard test in the notation of
this paper.  $\bar \chi^2$ is constructed in two steps.  First one
picks a time interval $[t_1, t_1+\tau]$ whose length $\tau$ is not
less than the time duration of the signal-model template $T$ convolved
with $S_n^{-1}$, and which includes the support of $S_n^{-1} * T$
\footnote{The choice of time interval is not explicitly discussed in
\cite{baggioetal} but the natural choice is the support of $S_n^{-1} *
T$}.

Then one constructs a function of a single real amplitude $A$ (or,
potentially, additional parameters describing the signal)
\begin{equation}
\label{e:vitale}
\bar \chi^2 (A) = \bigl( \tilde s - A \tilde T, \tilde s - A \tilde T) =
\bigl( \tilde s , \tilde s ) - 2 A \bigl( \tilde s , \tilde T ) + A^2.
\end{equation}
Here, $\tilde s$ is computed from (\ref{e:fft1}), but the integral is
taken {\it only} over the time interval $[t_1, t_1+\tau]$.  As a
function of $A$, $\bar \chi^2$ has an absolute minimum at
\[
A = \bigl( \tilde s , \tilde T ) = z.
\]
The minimum value of $\bar \chi^2(A)$ defines $\bar \chi^2$:
\begin{equation}
\label{e:minbarchi2}
\bar \chi^2 = \bigl( \tilde s , \tilde s ) - \bigl( \tilde s , \tilde T )^2 =
\bigl( \tilde n , \tilde n ) - \bigl( \tilde n , \tilde T )^2.
\end{equation}
Thus, $\bar \chi^2$ measures the difference between the squared
amplitude of the detector output and the squared SNR.  It is clear from
this equation that $\bar \chi^2$ is quite different from the $\chi^2$
discriminator defined in this paper. In particular, if the detected
$SNR$ vanishes ($z=0$) then $\bar \chi^2 = (\tilde s, \tilde s)$,
whereas $\chi^2 = \sum_{j=1}^p \bigl( \tilde s , \tilde T )_j^2$.  In this
case $\bar \chi^2$ is measuring the ``total length'' of $\tilde s$,
while $\chi^2$ is measuring the sum of squares of the components of
$\tilde s$ obtained by projecting it onto $p$ orthonormal components
of $\tilde T$.

It is also instructive to compute the expected value of $\bar \chi^2$
in our frequency-domain-based formalism \footnote{Reference
\cite{baggioetal} assumes that the detector noise is stationary and
Gaussian, and then constructs a new basis (not an orthogonal
transformation of the old basis) which diagonalizes the inverse
correlation function $\mu_{ij}$.  Although the authors do not say so,
if the noise is stationary then $\mu_{ij}=f(i-j)$ is of Topelitz form
and depends only upon $i-j$.  One can then show that for large $N$ the
diagonal basis is the frequency basis obtained via a Discrete Fourier
Transform. So in effect \cite{baggioetal} {\it is} done in a
frequency basis.}. Denote the instrument's data acquisition sample time
by $\Delta t$, so that the Nyquist frequency is $f_N = 1/(2 \Delta t)$
and the number of data samples is $N=\tau/\Delta t$.  After setting
$s(t)$ to zero outside of the time interval $[t_1, t_1+\tau]$, one has
\begin{eqnarray}
\label{e:finitetime}
\nonumber
\langle |\tilde n(f) |^2 \rangle
& = & \int_0^\tau dt \int_0^\tau dt' \langle n(t+t_1) n(t'+t_1) \rangle {\rm e}^{2\pi i f(t-t')} \\
\nonumber
& = & \int df' \int_0^\tau dt \int_0^\tau dt' \; S_n(f') {\rm e}^{2\pi i (f-f')(t-t')} \\
\nonumber
& = & \int df' \int_0^\tau dt \int_{-\infty}^\infty dt' \; S_n(f') {\rm e}^{2\pi i (f-f')(t-t')} \\
\nonumber
& = & \int df' \int_0^\tau dt \; S_n(f') \delta(f-f') {\rm e}^{2\pi i (f-f')t} \\
& = &  \int_0^\tau dt \; S_n(f) = \tau S_n(f).
\end{eqnarray}
In going from the second to the third line, we have assumed that
$\tau$ is greater than the characteristic time over which the
autocorrelation function of the noise falls off.  From
(\ref{e:finitetime}) it follows immediately that
\begin{equation}
\label{e:finitetime2}
\langle (\tilde n, \tilde n) \rangle = \int_{-f_N}^{f_N} 
{\langle | \tilde n(f)|^2 \rangle \over S_n(f)} df = 2 f_N \tau = N,
\end{equation}
where, as before, $N$ is the number of data samples.  And provided
that the interval $[t_1, t_1+\tau]$ includes the support of the
template $T$, we have already shown that $\langle (\tilde n, \tilde
T)^2 \rangle = 1$.  Combining this with (\ref{e:minbarchi2}) and
(\ref{e:finitetime2}) one finds the expectation value
\begin{equation}
\langle \bar \chi^2 \rangle = N - 1,
\end{equation}
corresponding to the fact that $\bar \chi^2$ has a classical $\chi^2$
distribution with $N-1$ degrees of freedom.

As this analysis and counting makes clear, the definition of $\bar
\chi^2$ given in \cite{baggioetal} includes the degrees of freedom
associated with {\it every} pixel in the time-frequency plane.  In
contrast to this, the $\chi^2$ time-frequency discriminator defined in
this paper includes only blocks of pixels centered along the
time-frequency track of the template $T$.  In fact, when $\bar \chi^2$
is actually computed from data, the number of degrees of freedom is
reduced to include only those degrees of freedom in the sensitive band
of the detector.
\footnote{At the end of Section~V and the beginning of Section~VI of
\cite{baggioetal}, the authors describe how the signal was
down-sampled and frequency shifted to reduce the number of degrees of
freedom $N$.  This corresponds to choosing regions of the
time-frequency plane that have good overlap with the putative signal
and expected sources of electronic and instrumental noise.  It may be
possible to redefine $\bar \chi^2$ in a way that automatically
includes only the relevant regions of the time-frequency plane, and
makes it invariant under transformations such as oversampling that
should leave it invariant.}

\section{Conclusion}
\label{s:conclusion}

This paper defines a $\chi^2$ time-frequency discriminator which is an
effective veto for the output of a matched filter.  The statistic
looks along the time/frequency track of purported signal to see if the
SNR accumulates in a way that is consistent with the properties of the
signal and the second-order statistics of the detector's noise. Small
values of $\chi^2$ are consistent with the hypothesis that the
observed SNR arose from a detector output which was a linear
combination of Gaussian noise and the putative signal waveform.  Large
values of $\chi^2$ indicate that either the signal did not match the
template, or that the detector was producing very non-Gaussian
noise. The method appears to work well for broadband detectors and
signals, and may have wider applicability.

The main results of the paper are the definitions of $\chi^2$ given in
(\ref{e:definechisquare}) and (\ref{e:twophasedef}), and equations
(\ref{e:mainresult1}), (\ref{e:mainresult2}), and
(\ref{e:mainresult3}) which give upper bounds on the expected value of
$\chi^2$ if the signal and template are slightly mis-matched.  We also
showed that the $\chi^2$ time-frequency discriminator is distinct from
the standard ``goodness of fit'' $\chi^2$ test described in
\cite{baggioetal}.

Recently the TAMA group has been experimenting with using
$|z|^2/\chi^2_{\rm r}$ as a thresholding statistic for detection
purposes \cite{tamainspiral4}, where $\chi^2_{\rm r}$ is $\chi^2$
divided by the number of degrees of freedom.  In Monte-Carlo
simulation studies, they have shown that this prevents simulated high
SNR events from being rejected by the discriminator. This is one way
to accommodate mismatch between templates and signals.

The construction of $\chi^2$ requires the (a-priori or posterior)
choice of how many frequency bands to use.  An
outstanding research question is ``what is the best way to set the
value of $p$?''  The correct answer to this question probably depends
upon a number of factors.  These include:
\begin{itemize}
\item
The ultimate goal of the analysis (i.e., setting upper limits, or
detecting sources).
\item
The statistical properties of the detector noise (both broadband
background and transient glitches).
\item
The maximum mismatch $\epsilon$ of the template bank.
\item
The accuracy to which the putative signal waveforms can be calculated
or predicted.
\end{itemize}
One possible answer comes from the behavior of $\chi^2$ as a function
of the template mismatch $\epsilon$.  We have shown that there are two
possible types of behavior, depending upon whether or not the two
templates have the same power spectrum (which implies that they share
the same intrinsic frequency bands).  In some situations, it may make
sense to work along the boundary in the $(p, \epsilon)$ plane that
separates these two types of behavior, as shown in
Figure~\ref{f:limits}.

Some interesting work on this topic has been done by Babak
\cite{Babak} who has found the optimal value of $p$ for the GEO
detector by studying the relative distributions of $\chi^2$ in the
presence and absence of simulated inspiral chirp signals.

A related issue concerns the construction of a template bank. The
minimum number of required templates is fixed by physics and the
behavior of the detector: one divides the volume of parameter space by
the volume covered per template \cite{Owen}. However within this
constraint the actual locations of the templates and their precise
parameters are quite arbitrary.  It may be possible to break this
degeneracy by constructing a template bank in such a way that the
effects on $\chi^2$ of a signal/template mismatch are minimized, or
bounded significantly {\it below} the absolute limits that we have
obtained. Roughly speaking this corresponds to placing the templates
in such a way that the overlap $(\tilde T, \tilde T')_j$ is
simultaneously maximized in each of the different bands $j=1, \cdots,
p$.  This might also require varying the value of $p$ as one moves
across the template bank.

While the $\chi^2$ test was specifically constructed for broadband
signals, it may be generalized to signals that are normally thought of
as ``narrow-band''.  One example is the Continuous Wave (CW) signals
expected from a rapidly rotating neutron star (pulsar).  In fact,
these CW signals are not so ``narrow-band''.  Typically, the Earth's
motion around the solar system modulates such a signal by a part in
$10^4$ over a six-month-long observation.  Since the intrinsic
frequency of such a source is of order 1~kHz, and the frequency
resolution during six months is of order $10^{-7}$~Hz, these signals
are actually spread over approximately $10^6$ frequency bins.  Thus a
$\chi^2$ test could be employed for such signals.

In fact a corresponding $\chi^2$ test could be implemented in the time
domain for any type of signal.  In effect, one simply breaks the
template (viewed as a function of time) into $p$ contiguous and
non-overlapping sections, each of which gives an equal expected
contribution to the total SNR \footnote{For signals like inspiral
chirp signals, which have monotonically increasing frequency as a
function of time, the different contiguous frequency bands $\Delta
f_j$ are in one-to-one and monotonic correspondence with the adjacent
regions of time.}. One then forms the $\chi^2$ statistic by seeing if
these relative contributions are clustered around the expected SNR
(which is a fraction $1/p$ of the total SNR).  Note that an analysis
like the one done in this paper shows that this quantity does {\it
not} have a classical $\chi^2$ distribution if the detector noise is
Gaussian and {\it colored}.  This is because the noise in two
non-overlapping time intervals is correlated.  However if the length
of the time intervals is long compared to the characteristic
correlation time of the noise, or if the detector output and template
are whitened, then the resulting quantity would have a classical
$\chi^2$ distribution for Gaussian noise.

\acknowledgments
The author thanks Jolien Creighton for many useful conversations,
suggestions and corrections. He also thanks Kip Thorne for
interactions with his research group that stimulated much of the early
work on this topic. Stanislav Babak, Jolien Creighton, Albert
Lazzarini, and Peter Shawhan helped to clarify the manuscript's
wording and pointed out many typographical errors.  Eric Key (UWM
Mathematics Department) provided a rigorous proof that
(\ref{e:convexresult}) is the maximum of $\kappa$ by showing that
$\kappa$ is a convex function on a convex set.

This research was supported in part by NSF grants PHY-9507740,
PHY-9728704, PHY-0071028, and PHY-0200852, and by the LIGO Visitors
Program.  Much of this work was done while the author was visiting the
LIGO laboratory and the TAPIR group at Caltech.

\appendix
\section{Distribution of $\chi^2$ for stationary Gaussian detector noise}
\label{s:appendix}

Here, we derive the probability distribution function (pdf) of the
$\chi^2$ discriminator under the assumption that the detector's noise
is stationary and Gaussian. For simplicity we treat the single-phase
case; the two-phase case corresponds to replacing $p$ and $p-1$ by
$2p$ and $2p-2$ respectively.

Since the different $z_j$ are each constructed from different,
non-overlapping frequency bands, they are themselves Gaussian random
variables.  Hence their pdf is
\begin{equation}
\label{e:prob1}
P(z_1,\cdots,z_p) = \prod_{j=1}^p (2 \pi \sigma)^{-1/2} 
{\rm e}^{- { \left[z_j - \alpha /p \right]^2 / 2 \sigma}}
\end{equation}
with $\sigma = 1/p$ and $\alpha = \langle z \rangle$.

We need to calculate the pdf of $\Delta z_j = z_j - z/p$.  This is
complicated by the fact that these variables are correlated \footnote{
In an early version of the GRASP manual, this correlation was
neglected and a $\chi^2$ distribution with $p$ rather than $p-1$
degrees of freedom was obtained.  This mistake was pointed out by
Jolien Creighton.} since their sum vanishes exactly. We denote the pdf
of $\Delta z_j$ by $\bar P(\Delta z_1,\cdots,\Delta z_p)$.  It is
defined by the relation that the integral of any function of $p$
variables $F(u_1,\cdots,u_p)$ with respect to the measure defined by
this probability distribution satisfies
\begin{eqnarray}
\nonumber
 \int  du_1 & \cdots & \int du_p  \bar P (u_1,\cdots,u_p)  F(u_1,\cdots,u_p) = \\
\label{e:defpbar}
 \int dv_1 & \cdots & \int dv_p   P (v_1,\cdots,v_p) \times \\
\nonumber
 & & \quad F(v_1 -  \sum_{j=1}^p { v_j \over p} ,\cdots,v_p - \sum_{k=1}^p {v_k \over p} ). 
\end{eqnarray}
We can use this definition to find a closed form expression for $\bar
P$.

Let $F(u_1,\cdots,u_p) = \prod_{j=1}^p \delta(u_j - \Delta z_j)$ in
(\ref{e:defpbar}).  One obtains
\begin{eqnarray}
&& \bar P (\Delta z_1,\cdots,\Delta z_p) = \\
\nonumber
&&  \prod_{j=1}^p \int  dv_j {
{\rm e}^{-[v_j - \alpha/p]^2 / 2 \sigma} \over
(2 \pi \sigma)^{1/2}} \; \delta(v_j -
\Delta z_j - \sum_{j=1}^p {v_j \over p}).
\end{eqnarray}
To evaluate the integral, change to new variables $w_1, \cdots,
w_{p-1}, W$ defined by
\begin{eqnarray}
\nonumber
v_1 & = &W/p +  w_1 \\
\nonumber
& \cdots & \\
v_{p-1} & = & W/p + w_{p-1} \\
\nonumber
v_p & = &  W/p - w_1 - \cdots - w_{p-1}.
\end{eqnarray}
The Jacobian of this coordinate transformation is
\begin{eqnarray}
\nonumber
J & = & \det \left[ { \partial(v_1,\cdots,v_p ) \over \partial(w_1,\cdots,w_{p-1},W) } \right] \\
& = & \det \left[ \matrix{
                   1 & 0 & \cdots &  0 & 1/p \cr
                   0 & 1 & \cdots &  0 & 1/p \cr
                     &   & \cdots &    &     \cr
                   0 & 0 & \cdots &  1 & 1/p \cr
                  -1 &-1 & \cdots & -1 & 1/p \cr
} \right].
\end{eqnarray}
Using the linearity in rows of the determinant, it is straightforward to show that $J=1$.

The integral may now be written as
\begin{eqnarray}
\nonumber
&& \bar P (\Delta z_1,\cdots,\Delta z_p) = \int  dw_1 \cdots \int dw_{p-1} \int dW \times \\
\nonumber
&& \quad (2 \pi \sigma)^{-p/2} {\rm e}^{-[(v_1 - \alpha/p)^2 + \cdots + (v_p - \alpha/p)^2]/ 2 \sigma} \times  \\
\nonumber
&& \quad \delta(w_1-\Delta z_1)  \cdots  \delta(w_{p-1}-\Delta z_{p-1}) \times \\
&& \quad \delta(w_1+\cdots + w_{p-1}+\Delta z_p).
\end{eqnarray}
The argument of the exponential may be expressed in terms of the new
integration variables as
\begin{eqnarray}
&& (v_1 - \alpha/p)^2 + \cdots + (v_p - \alpha/p)^2 = \\
\nonumber
&& w_1^2 + \cdots + w_{p-1}^2 + (W-\alpha)^2/p + (w_1 + \cdots + w_{p-1})^2
\end{eqnarray}
and thus the integral yields
\begin{eqnarray}
\label{e:deltaprob}
\nonumber
&& \bar P (\Delta z_1,\cdots,\Delta z_p)  \\
\nonumber
&& = \int dW (2 \pi \sigma)^{-p/2} {\rm e}^{-[\Delta z_1^2 + \cdots + \Delta z_p^2 + (W-\alpha)^2/p]/2\sigma} \times \\
\nonumber
&& \qquad \delta(\Delta z_1 + \cdots +\Delta z_p) \\
\nonumber
&& = (2 \pi \sigma)^{-p/2} (2 \pi \sigma p)^{1/2} {\rm e}^{-[\Delta z_1^2 + \cdots + \Delta z_p^2]/2 \sigma} \times \\
&& \qquad \delta(\Delta z_1 + \cdots +\Delta z_p).
\end{eqnarray}
This pdf is easily visualized.  In $\mathbb{R}^p$ it vanishes except on the
$(p-1)$-plane $\Delta z_1+ \cdots \Delta z_p=0$.  On that hyperplane
it is a spherically symmetric function of the distance from the origin.

This probability distribution arises because we do not know the true
expectation value $\langle z \rangle$ but can only estimate it using
the single measured value of $z$.  This issue arises whenever the mean
of a distribution is not know but must be estimated (problem 14-7 of
\cite{matthewsandwalker}).  This probability distribution is ``as
close as possible to a Gaussian" subject to the constraint that the
sum of the $\Delta S_j$ must vanish.  It is significant that this pdf
is completely independent of $\alpha$, which means that if the
detector noise is Gaussian then the properties of the $\Delta z_j$ do
not depend upon whether a signal is present or not.

We can now compute the probability distribution of $\chi^2 = p (\Delta
z_1^2 + \cdots \Delta z_p^2)$ using (\ref{e:deltaprob}).  The
probability that $\chi^2 < \chi_0^2 $ is the integral of
(\ref{e:deltaprob}) inside a sphere of radius $\chi_0/\sqrt{p}$. To
evaluate the integral, introduce a new set of coordinates
$(x_1,\cdots,x_p)$ on $\mathbb{R}^p$ obtained from any special
orthogonal $SO(p)$ transformation of the $p$ coordinates $\sqrt{p}
(\Delta z_1,\cdots,\Delta z_p)$ for which the {\it new} $p$'th
coordinate is orthogonal to the hyperplane $\Delta z_1 + \cdots
+\Delta z_p = 0$.  For example take $x_p = \Delta z_1 + \cdots +\Delta
z_p$.  Let $r$ denote the radius from the origin $r^2 = x_1^2 + \cdots
+ x_p^2$, and note that $\chi^2 = r^2$.  The probability is then
\begin{eqnarray}
\nonumber
P(\chi < \chi_0  ) & = & \int_{r < \chi_0 }  \bar P (\Delta z_1,..., \Delta z_p)\;  p^{-p/2} \; d^p x\\
\nonumber
 & = &   (2 \pi \sigma p)^{{1 \over 2}-{p \over 2}} \int_{r < \chi_0} 
     {\rm e}^{-r^2/2} \delta(x_p) \; d^p x.
\end{eqnarray}
The integral over the coordinate $x_p$ is trivial, yielding a
spherically-symmetric integral over $\mathbb{R}^{p-1}$:

\begin{equation}
P(\chi < \chi_0)  =  (2 \pi \sigma p)^{{1 \over 2}-{p \over 2}} 
\int_{ r < \chi_0   } 
     {\rm e}^{-r^2/2}  \; d^{\, p-1}x.
\end{equation}
Since this is spherically symmetric, we can write the volume element $
d^{\; p-1}x = \Omega_{p-2} r^{p-2} dr$ where $\Omega_n = {2
\pi^{(n+1)/2} \over \Gamma({n+1 \over 2})}$ is the $n-$volume of the
unit-radius $n-$sphere $S^n$. One then has
\[
P(\chi < \chi_0) = (2 \pi \sigma p)^{{1 \over 2}-{p \over 2}} \Omega_{p-2}
\int_0^{\chi_0} r^{p-2} {\rm e}^{-r^2/2} dr.
\]
Changing variables to $u = r^2/2$ this takes the form
\begin{eqnarray}
\nonumber
P(\chi < \chi_0) & = & {1 \over \Gamma({p \over 2} - {1 \over 2})} \int_0^{\chi_0^2/2 }
  u^{{p \over 2}-{3 \over 2}} {\rm e}^{-u} du \\
\nonumber
& = &  {\gamma({p \over 2}-{1
\over 2}, { \chi_0^2 \over 2}) \over \Gamma({p \over 2}-{1 \over 2})}
\end{eqnarray}
which is the classical $\chi^2$ cumulative distribution for $p-1$ real
degrees of freedom, expressed in terms of the incomplete
$\gamma$-function.

\section{Max of A cos($\psi$) + B sin($\psi$)}
\label{s:max}
Twice in this paper, we require the maximum of $f(\psi) = A \cos \psi
+ B \sin \psi$, for fixed values of $A$ and $B$, as $\psi$ varies in
the interval $[0,2\pi)$.  This is trivial to obtain.  Setting the
derivative $df/d\psi$ to zero gives $ B \cos \psi_0 -A \sin \psi_0 =
0$, implying that at the maximum $\tan \psi_0 = B/A$.  Thus one has
$\sec^2 \psi_0 = 1+\tan^2 \psi_0 = 1+ B^2/A^2$ and hence $\cos \psi_0
= A/\sqrt{A^2+B^2}$ and $\sin \psi_0 = B/\sqrt{A^2+B^2}$.
Substituting these into $f$ yields the maximum value
$f(\psi_0)=\sqrt{A^2+B^2}$.



\vfill
\end{document}